\documentclass[aps,pre,10pt,twocolumn,superscriptaddress,nofootinbib]{revtex4-2}

\usepackage{amsmath,amssymb,times,hyperref,graphicx}
\hypersetup{
    colorlinks=true,       
    linkcolor=blue,          
    citecolor=blue,        
    filecolor=blue,      
    urlcolor=blue          
}

\newcommand{\s}{\mathrm{s}}
\newcommand{\ns}{\mathrm{ns}}
\renewcommand{\d}{\mathrm{d}}
\newcommand{\g}{\mathrm{g}}
\renewcommand{\L}{\mathrm{L}}
\renewcommand{\l}{\mathrm{l}}
\newcommand{\kt}{k_\mathrm{B}T}

\begin{document} 

\title{Accelerated Ostwald ripening by chemical activity} 

\author{Benjamin Sorkin}
\email{bs4171@princeton.edu}
\affiliation{Princeton Center for Theoretical Science, Princeton University, Princeton, NJ 08544, USA}

\author{Ned S. Wingreen}
\email{wingreen@princeton.edu}
\affiliation{Princeton Center for Theoretical Science, Princeton University, Princeton, NJ 08544, USA}
\affiliation{Lewis-Sigler Institute for Integrative Genomics, Princeton University, Princeton, NJ 08544, USA}
\affiliation{Department of Molecular Biology, Princeton University, Princeton, NJ 08544, USA}

\begin{abstract}
    Phase separation of biomolecular condensates promotes membrane-free compartmentalization in cells. The dynamics of these biocondensates is routinely regulated  by energy-consuming processes. Here, we devise a theory pinpointing how active chemical reactions, interconverting molecules between phase-separating and inert forms, can drive faster condensate coarsening. We find that mass conservation limits droplet volume growth to being linear in time regardless of activity, resembling the passive Lifshitz-Slyozov law. However, if reactions are restricted to occur only outside droplets, the rate of Ostwald ripening can be increased by an arbitrarily large factor. Our theory is quantitatively supported by recent experiments on ripening in the presence of fueled interconversion reactions, under precisely the predicted conditions. We posit that the ability to induce rapid biocondensate coarsening can be advantageous in synthetic-biological contexts, \textit{e.g.}, as a regulator of metabolic channeling.
\end{abstract}

\maketitle

Ostwald ripening refers to the thermodynamic coarsening process of solute-rich phase-separated droplets as a result of surface tension. Smaller droplets shrink due to their larger Laplace pressure whereas larger droplet grow via the diffusion of solute. Average droplet size may also increase due to coalescence events. These processes continue in principle until phase-separation completes. However, a vastly richer physics emerges when activity and energy consumption are introduced. One well-studied outcome is ``size control''~\cite{ZwickerARXIV25,WeberROPP19,ZwickerPRE15,ZwickerNP17,SastreNC25}, whereby active processes prevent unbounded coarsening. Size control can be achieved via nonequilibrium reactions that interconvert the solute between phase-separating and inert forms. Additional possible outcomes of activity include arrested ripening~\cite{BauermannARXIV24} (where small droplets do not shrink), emergence of Turing patterns~\cite{CaratiPRE97,AslyamovPRL23} (where the dense phase is destabilized by the coupling between thermodynamics and reactions), and reversal of Ostwald ripening~\cite{TjhungPRX18} (where dilute bubbles form inside the dense phase). Further complexities such as elasticity~\cite{RosowskiNP20,FernandezNM24,MannattilARXIV24,QiangPRX24} or charge asymmetry~\cite{LuoARXIV25} extend the list of possible outcomes. In contrast to these effects which oppose Ostwald ripening, it was recently demonstrated experimentally that fueled interconversion reactions in organochemical emulsions can subtantially accelerate ripening~\cite{TenaSolsonaCSC21}. How do these same interconversion reactions lead to such a different outcome?

One motivation to investigate accelerated Ostwald ripening is its potential biological relevance. Phase separation of biomolecules to form biomolecular condensates (also termed ``membraneless organelles'') is pervasive in cells, enabling the compartmentalization of biochemical processes~\cite{BananiNRCB17,ShinSCI17}. Biocondensate size can be crucial for function~\cite{LeeNP2023,MartinezCalvoARXIV24,BuchnerJCP13,CastellanaNB14,ZhaoNCB19}.
For example, consider the algal pyrenoid~\cite{ShanPLANT23}\,---\,a condensate composed of Rubisco (a carbon-fixating enzyme) and EPYC1 (a ``sticky'' linker protein that drives phase separation)\,---\,which is dynamically regulated through energy-consuming phosphorylation cycles. During cell division, phosphorylation renders EPYC1 less sticky, breaking the pyrenoid into small droplets via a size-control mechanism, thus favoring equipartition of Rubisco between the daughter cells. After division, dephosphorylation allows recoarsening into a single pyrenoid in each daughter cell. For cases like the pyrenoid where a single condensate is advantageous, could cells use energy to accelerate Ostwald ripening, and how might this be achieved?

\begin{figure}[t!]
    \centering
    \includegraphics[width=0.99\linewidth]{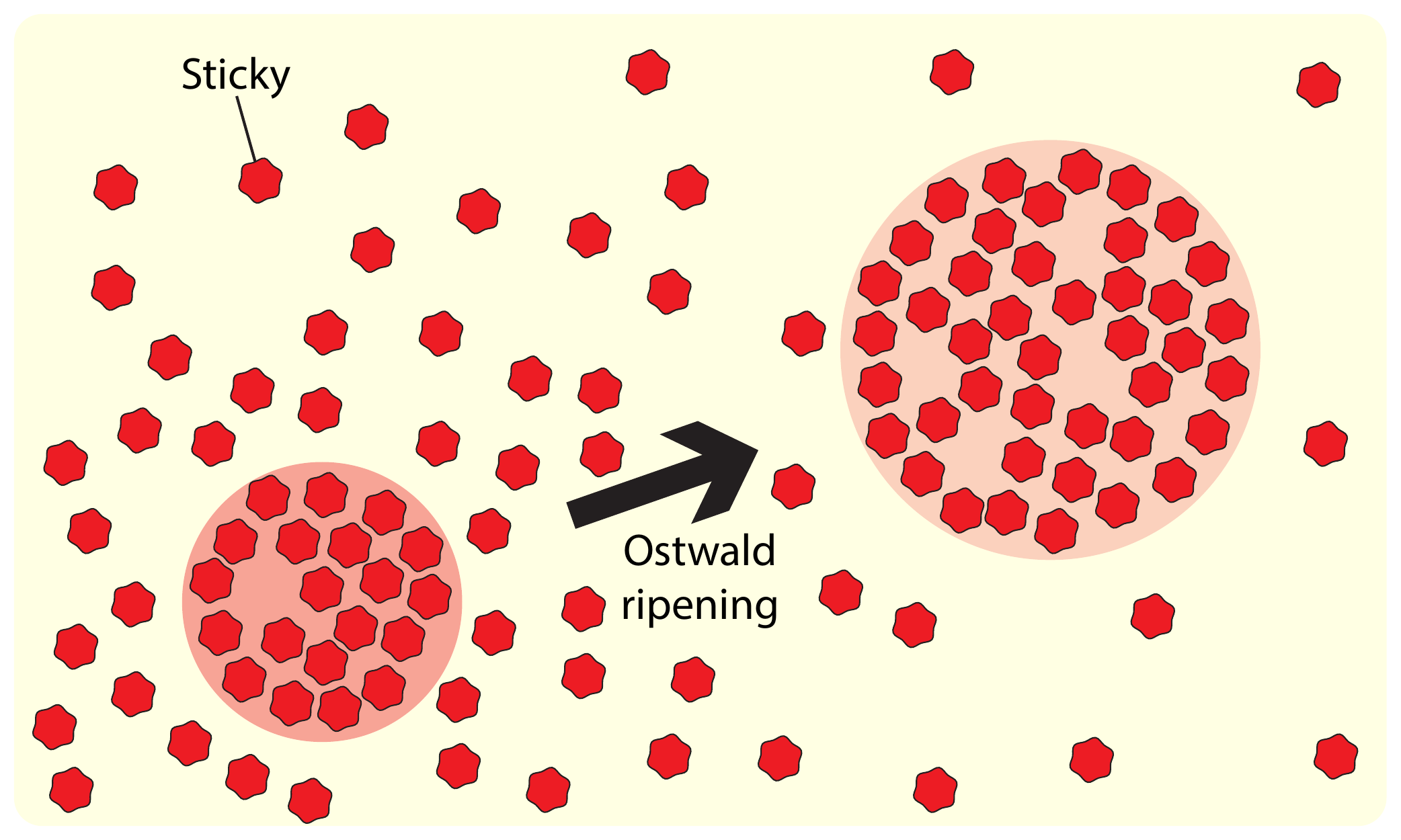}
    \includegraphics[width=0.99\linewidth]{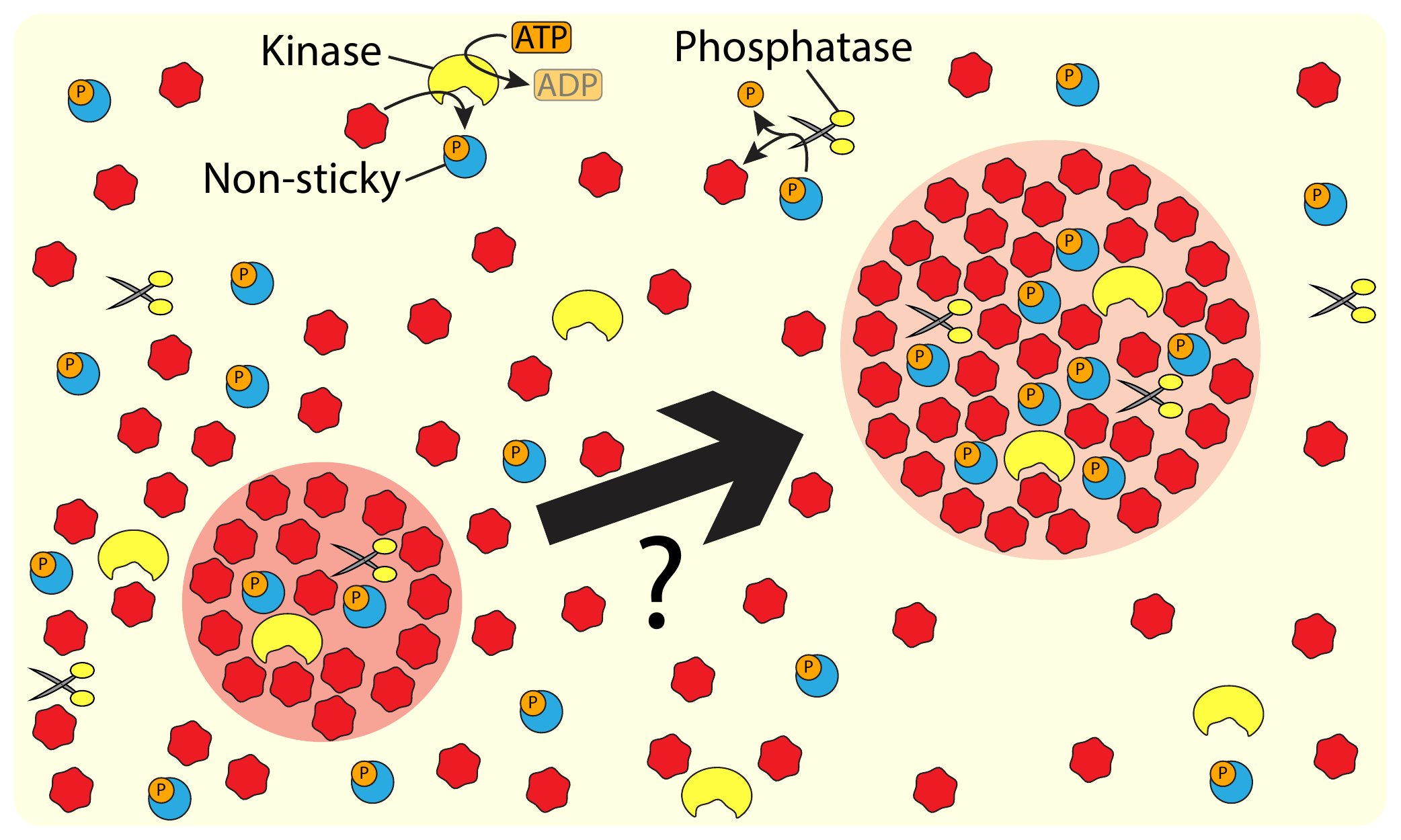}
    \caption{Illustration of the model. (a) Passive system: A supersaturated monodisperse suspension of ``sticky'' particles. Spontaneously-formed dense-phase nuclei will coarsen at the expense of the material available in the surrounding dilute phase. Once the supersaturation becomes small, Ostwald ripening commences\,---\,larger droplets keep growing while smaller droplets, whose Laplace pressure is higher, begin shrinking. (b) Active system: Motivated by the active ripening of the algal pyrenoid~\cite{ShanPLANT23}, we introduce kinases and phosphatases to the above suspension. The former phosphorylate the sticky proteins, rendering them ``nonsticky'' (inert), and the latter dephosphorylate nonsticky proteins returning them to the sticky form. In our study, we investigate whether these active processes can allow for faster Ostwald ripening.}
    \label{fig:illustration}
\end{figure}

To answer this question, we consider a solution consisting of a ``sticky'' species that tends to phase separate, into which we introduce interconversion reactions that give rise to inert ``nonsticky'' species. Combining a reaction-diffusion approach~\cite{ZwickerARXIV25} with Lifshitz-Slyozov theory~\cite{LifshitzJPCS61}, we devise a general theory that pinpoints the requirements to achieve acceleration. Put simply, reactions in the dilute phase promote the coarsening flux of sticky particles, whereas conversions inside the droplets oppose their growth. In line with experiments~\cite{TenaSolsonaCSC21}, we detect that when the reactions inside droplets are negligible, acceleration of ripening by an arbitrary factor is possible. We note however that, mass conservation renders unrealizable any time scaling faster than the Lifshitz-Slyozov law, so only the rate coefficient can be tuned by activity.

Consider a solution of ``sticky'' particles (\textit{e.g.}, proteins) which tend to phase separate. They are immersed in a solvent (\textit{e.g.}, the cytoplasm, implicitly taking into account all other macromolecules of the cell) whose volume $V$ and temperature $T$ are fixed. We denote the local concentration of the sticky species as $\phi_\s$. Phase separation occurs whenever the solution's free energy has concave regions in the thermodynamic parameter space. We denote the coexistence concentration of the dense phase (sticky-rich, ``liquid'') by $\phi_{\s\l}^\circ$ and of the dilute phase (sticky-poor, ``gas'') by $\phi_{\s\g}^\circ$. When two contacting phases are in equilibrium, the two concentrations must be such that chemical potentials and pressures of the two phases equate~\cite{DoiBOOK13}. In the strong phase separation limit, setting $\phi_{\s\l}^\circ=1$, we expect $\phi_{\s\g}^\circ\ll1$. 

Ostwald ripening is driven by surface tension. Thus, we distinguish between the above ``bulk'' coexistence concentrations, $\{\phi_{\s\l}^\circ,\phi_{\s\g}^\circ\}$, and the concentrations found around droplets of finite radius $R$, $\{\phi_{\s\l}^\L(R),\phi_{\s\g}^\L(R)\}$. These concentrations differ since the liquid pressure must oppose both the gas pressure and so-called Laplace pressure, the latter resulting from droplet curvature~\cite{deGennesBOOK03}. For large droplets compared to interface width, the concentrations are connected via an Ostwald-Freundlich-type equation $\phi_{\s\alpha}^\L(R)=\phi_{\s\alpha}^\circ+\lambda_{\s\alpha}/R$ (with $\alpha=\l,\g$), where the $\lambda_{\s\alpha}>0$ are Laplace-pressure correction coefficients, proportional to surface tension and the $\alpha$th phase isothermal compressibility. 

To enable phase separation, the system is prepared in some initial, slightly supersaturated concentration denoted $\varphi^\infty\gtrsim\phi_{\s\g}^\circ$. This so-far passive solution is known to undergo Ostwald ripening; \textit{cf.} Fig.~\ref{fig:illustration}(a). As the saturation decreases during droplet growth, since a higher dilute-phase concentration $\phi_{\s\g}^\L(R)$ surrounds smaller droplets due to Laplace pressure, a concentration gradient diffusively transports material from small to large droplets. This is summarized by the radius growth rate equation for a focal droplet~\cite{LifshitzJPCS61},
\begin{equation}
    \frac{\d R}{\d t}=\frac{D_\g}{R}\left[(\varphi^\infty-\phi_{\s\g}^\circ)-\frac{\lambda_{\s\g}} R\right],\label{eq:dRdt_passive}
\end{equation}
where $D_\g$ is the dilute-phase diffusion constant
. Upon inspecting the size distribution of thus growing and shrinking droplets, the Lifshitz-Slyozov scaling law emerges\cite{LifshitzJPCS61},
\begin{equation}
    \langle R(t)\rangle^3-\langle R(0)\rangle^3=\frac49D_\g\lambda_{\s\g}t.\label{eq:passive_ripening}
\end{equation}
Are active sticky-to-nonsticky interconversion reactions capable of accelerating this growth law?

To a sticky-only solution of (super-) saturation $\varphi^\infty$, we add a catalyst (\textit{e.g.}, a kinase) that converts sticky proteins into nonsticky ones, and vice versa by another catalyst (\textit{e.g.}, a phosphatase); \textit{cf.} Fig.~\ref{fig:illustration}(b). For example, in an aqueous solution, the sticky and nonsticky species might be interconvertible hydrophobic and hydrophilic compounds. Since these reactions are independent of the underlying thermodynamics of phase separation, they normally consume fuel whose concentration we assume is fixed~\cite{ZwickerARXIV25,TenaSolsonaCSC21}. We define the total reaction rate, $s(\phi_\s,\phi_\ns)$, as a function of the local particle concentrations, with the convention that $s>0$ implies net conversion of sticky molecules to nonsticky. For example, a purely first-order reaction mechanism takes the form $s(\phi_\s,\phi_\ns)=k_{\s\to\ns}\phi_\s-k_{\ns\to\s}\phi_\ns$, where $k_{i\to j}$ are the conversion rates, with associated stoichiometric coefficients $\upsilon_\s=-1$ and $\upsilon_\ns=1$. 

By nonsticky we mean that these particles are inert, meaning that phase separation is still driven by the remaining sticky species. We denote the coexistence concentrations of the dense phase (sticky-rich, ``liquid'') by $\{\phi_{\s\l}^\L,\phi_{\ns\l}^\L\}$ and of the dilute phase (sticky-poor, ``gas'') by $\{\phi_{\s\g}^\L,\phi_{\ns\g}^\L\}$ around droplets. When the two phases in contact are in equilibrium, we now require equal sticky chemical potentials, nonsticky chemical potentials, and pressures. 

In accordance with the Gibbs phase rule~\cite{AtkinsBOOK23}, these three conditions are insufficient to determine all four concentrations. Therefore, we next impose material conservation. During coarsening, the net amount of sticky material joining the droplet must track with the net amount of nonsticky material being expelled~\cite{KuehmannMMTA96}. Explicitly,
\begin{eqnarray}
	\frac{\d R}{\d t}=\frac{J_{\s\l}(R)-J_{\s\g}(R)}{\phi_{\s\l}^\L(R)-\phi_{\s\g}^\L(R)}
    =\frac{J_{\ns\l}(R)-J_{\ns\g}(R)}{\phi_{\ns\l}^\L(R)-\phi_{\ns\g}^\L(R)},\label{eq:dRdt_flux}
\end{eqnarray}
where $J_{i\alpha}$ are the outward flux densities of the $i$th species in the radial direction within the $\alpha$th phase at the interface. In the strong phase separation limit of sticky species, the concentrations $\phi_{\s\l}^\L\simeq1$ and $\phi_{\s\g}^\circ\lesssim\phi_{\s\g}^\L\ll1$ remain unchanged despite the presence of the nonsticky species, while in view of the above, $\phi_{\ns\l}^\L\simeq0$, and so only $\phi_{\ns\g}^\L\ll1$ is unknown; see Sec.~SI.

The fluxes can be computed using rigorous approximations. Specifically, within linear-irreversible thermodynamics~\cite{deGrootBOOK84}, the concentrations respond linearly to gradients in the chemical potentials, which are determined by the concentration profiles. These concentrations, in turn, are affected by the active interconversion reactions occurring in the solution. We summarize below the derivation and its results, and defer all rigorous calculations to Sec.~SII. In what follows, we assume that droplet growth is quasistatic and that the activity does not modify surface thermodynamics~\cite{WeyerARXIV24,ChoJCP23}.

In the dilute phase, we suppose that the free energy assumes the form of an ideal gas, so Fickian diffusion applies. Likewise, we approximate the reactions as effectively first-order, $s(\phi_{\s\g},\phi_{\ns\g})\simeq k_{\s\to\ns}\phi_{\s\g}-k_{\ns\to\s}\phi_{\ns\g}$, as only the lowest-order reaction mechanisms take place with a nonzero rate in the small concentration limit. (We generalize this in Sec.~SII~D.) The effective rate constants $k_{\s\to\ns}$ and $k_{\ns\to\s}$ can be derived from the full reaction mechanism $s(\phi_{\s},\phi_{\ns})$, and they depend on fuel and catalyst concentrations. Due to the presence of reactions, we note that the total superaturation $\varphi^\infty$ is now redistributed among sticky and nonsticky species according to chemical equilibrium,
\begin{equation}
    \varphi_\s^\infty=\frac{k_{\ns\to\s}\varphi^\infty}{k_{\s\to\ns}+k_{\ns\to\s}}\text{\,\,\,and\,\,\,}\varphi_\ns^\infty=\frac{k_{\s\to\ns}\varphi^\infty}{k_{\s\to\ns}+k_{\ns\to\s}}.\label{eq:supersatur}
\end{equation}
Does this loss of sticky-particle saturation, $\varphi_\s^\infty<\varphi^\infty$, still allow for
accelerated Ostwald ripening?

\begin{figure}[t!]
    \centering
    \includegraphics[width=0.99\linewidth]{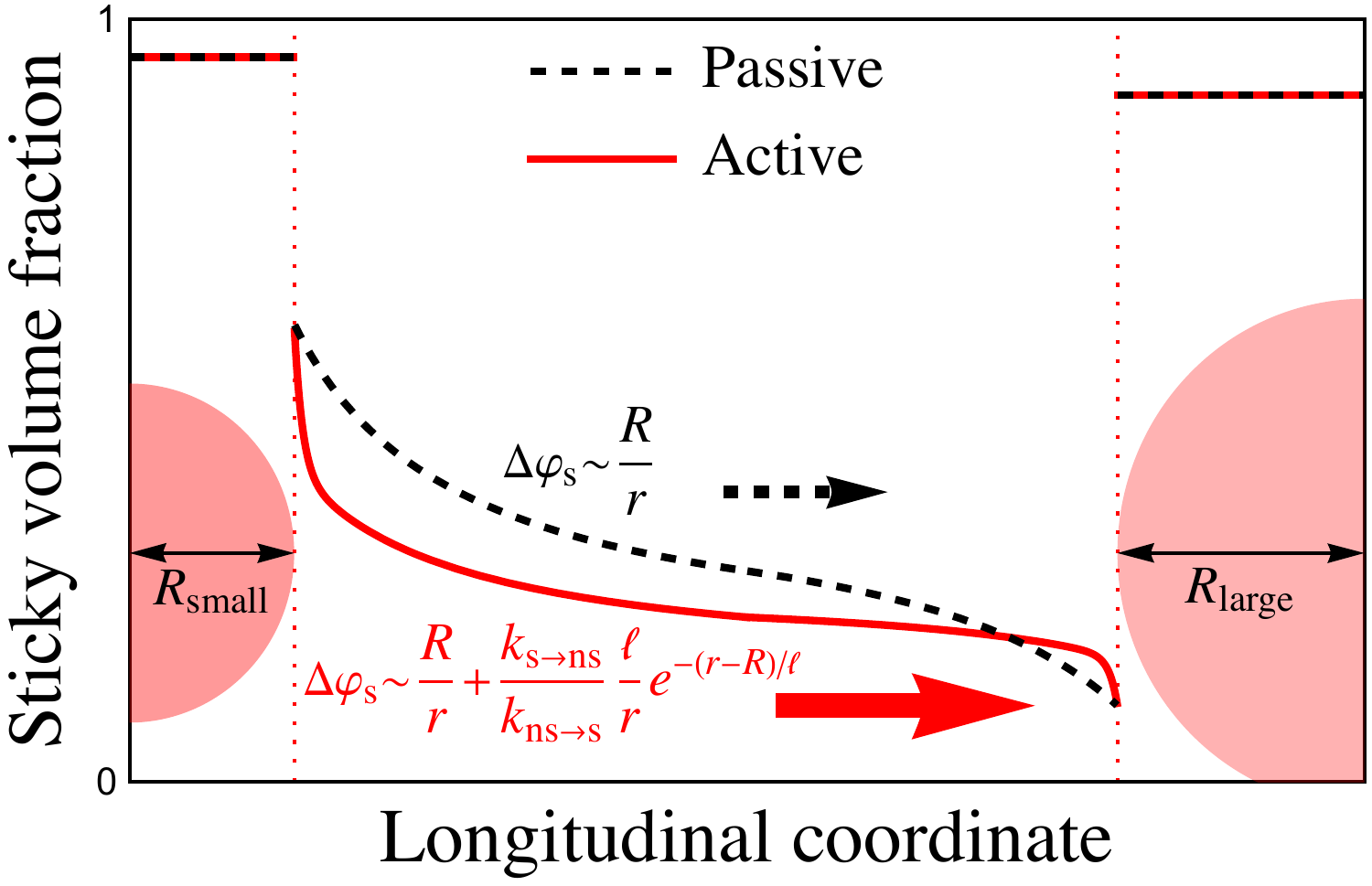}
    \caption{Illustrative concentration profiles between a small and a large droplet in the passive and active cases, where $\Delta\varphi_\s(r)=\phi_\s(r)-\phi_{\s\g}^\L$ is the excess saturation of sticky particles in the dilute phase. In the passive case, close to each of the droplets, the concentration profile is hyperbolic (dashed black curve), bringing material away from the small droplet and to the large one. In the active case, reactions create nonsticky material, thus overall lowering the sticky concentration in the dilute bulk. However, the reactions make the sticky profile steeper in the vicinity of droplets, thus increasing the flux of material from the small to the large droplet.  In both cases, the indicated proportionalities for $\Delta\varphi_\s$ hold for both the smaller and larger droplet, up to a change of sign for the large droplet. The full expressions for the concentration profiles can be found in Sec.~SII~F.}
    \label{fig:profiles}
\end{figure}

The competition between diffusion and reactions in the dilute phase introduces a lengthscale reflecting the diffusive distance over which reactions occur,
\begin{equation}
    \ell_\g=\sqrt{\frac{D_\g}{k_{\s\to\ns}+k_{\ns\to\s}}}.\label{eq:ellg}
\end{equation}
With its aid, upon finding the dilute-phase steady profiles from the reaction-diffusion equation, the material fluxes read
\begin{align}
    J_{\s\g}(R)=&-\frac{D_\g}R[\varphi_\s^\infty-\phi_{\s\g}^\L(R)]\nonumber\\&-\frac{D_\g}{\ell_\g}\frac{k_{\ns\to\s}\phi_{\ns\g}^\L(R)-k_{\s\to\ns}\phi_{\s\g}^\L(R)}{k_{\s\to\ns}+k_{\ns\to\s}},\label{eq:J_g}
\end{align}
where $J_{\ns\g}(R)$ is also given by Eq.~\eqref{eq:J_g} upon replacing $\s\leftrightarrow\ns$. 
The first term on the right side of Eq.~\eqref{eq:J_g} represents passive diffusion due to the concentration gradient from $\phi_{\s\g}^\L$ at the interface to $\varphi_\s^\infty$ in the bulk. The second term due to active reactions steepens the concentration gradient around the interface for both growing and shrinking droplets. Therefore, the effect of reactions is to further promote ripening. This fact is illustrated in Fig.~\ref{fig:profiles} and serves as the basis for the acceleration we report.

The dense phase is almost entirely composed of sticky proteins. This means that reactions inside the droplet can only hinder its growth, as their net effect is to convert sticky material into the nonsticky form, $s_\l\equiv s(\phi_{\s\l}^\L,\phi_{\ns\l}^\L)>0$. When the droplets are still small, the total amount of converted material per unit time is proportional to droplet volume. However, once the droplet is larger than $\ell_\l$\,---\,the distance a particle diffuses in the dense phase between reactions (analogous to $\ell_\g$, Eq.~\eqref{eq:ellg})\,---\,the reactions equilibrate at the core of the droplet, so the net loss of sticky material occurs only at its interface and hence the total conversion rate is proportional to droplet surface area. Thus, the dense-phase fluxes are the net reaction rate over droplet area,
\begin{equation}
    J_{\s\l}(R)=-J_{\ns\l}(R)=\begin{cases}
-\frac{1}{3}s_\l R, & R\ll\ell_\l,\\
-s_\l\ell_\l\left(1-\frac{\ell_\l}{\ell_\g}\frac{\ell_\g}R\right), & R\gg\ell_{\l}.
\end{cases}\label{eq:J_l}
\end{equation}
By contrast, Ostwald ripening, which opposes this loss of material, occurs with a rate proportional to droplet radius; the gas-phase reactions contribute additional ripening at a rate proportional to the area as well.

With these preparations, we manipulate Eq.~\eqref{eq:dRdt_flux} into the form
\begin{equation}
    \frac{\d R}{\d t}=\frac{D_\g}{R}[\varphi^\infty-\phi_{\s\g}^\L(R)-\phi_{\ns\g}^\L(R)].\label{eq:dRdt_cons}
\end{equation}
Eq.~\eqref{eq:dRdt_cons} is a cruical outcome\,---\,upon comparison with Eq.~\eqref{eq:dRdt_passive}, it emphasizes that for arbitrary activity, this system is bound by the scaling laws of conventional Ostwald ripening. This must be so due to the existence of a conserved concentration field; indeed the total concentration $\phi_\s+\phi_\ns$ is unaffected by activity. Therefore, in our search for acceleration, we learn that the scaling laws expected from Ostwald ripening are the best case scenario, requiring small $\phi_{\ns\g}^\L$, and our only hope is to increase the rate coefficient. Otherwise, in the ``worst case scenario'' of large $\phi_{\ns\g}^\L$, Ostwald ripening can be prevented entirely. 

From the second equality of Eq.~\eqref{eq:dRdt_flux} (the material conservation constraint~\cite{KuehmannMMTA96}), we obtain the remaining unknown $\phi_{\ns\g}^\L(R)$ in terms of a set of unitless parameters. Inserting this $\phi_{\ns\g}^\L$ in Eq.~\eqref{eq:dRdt_cons}, we find the explicit radius growth rate equation akin to Eq.~\eqref{eq:dRdt_passive}, amended to the active case. The values of these unitless parameters determine whether the system achieves acceleration. To obtain acceleration, we require $s_\l\ell_\l^2/D_\g\ll(\varphi_\s^\infty -\phi_{\s\g}^\L)$ and $s_\l\ell_\l^2/D_\g\ll(1+\ell_\g/\ell_\l)(\lambda_{\s\g}/\ell_\g)$. Otherwise, size control~\cite{ZwickerPRE15} (where a finite, stable droplet radius is reached instead of unlimited growth) or arrested ripening~\cite{BauermannARXIV24} (no shrinkage of small droplets), respectively, occur instead of accelerated ripening; see details in Sec.~SII~F and Fig.~S1.
We must further assume that the interplay between dense-phase reactions and thermodynamics do not destabilize the dense phase in the first place~\cite{CaratiPRE97,AslyamovPRL23,TjhungPRX18}; see Sec.~SII~E.

Supposing both inequalities for the dense-phase activity $s_\l\ell_\l^2/D_\g$ hold and the dense phase is stable, we arrive at the active counterpart of Eq.~\eqref{eq:dRdt_passive},
\begin{equation}
	\frac{\d R}{\d t}=\frac{D_\mathrm{eff}}{R}\left[(\varphi_\mathrm{eff}^\infty-\phi_{\s\g}^\circ)-\frac{\lambda_{\s\g}}R\right].\label{eq:dRdt_active}
\end{equation}
Here, we identified the effective diffusion constant
\begin{equation}
	D_\mathrm{eff}=\left(1+\frac{k_{\s\to\ns}}{k_{\ns\to\s}}\right)D_\g,
\end{equation}
which is enhanced by the dilute-phase reactions, and the effective supersaturation,
\begin{equation}
	\varphi_\mathrm{eff}^\infty=\frac{k_{\ns\to\s}\varphi^\infty-s_\l\ell_\l/\ell_\g}{k_{\s\to\ns}+k_{\ns\to\s}}
\end{equation}
which is further reduced from the dilute-phase chemical-equilibrium value (Eq.~\eqref{eq:supersatur}) due to the dense-phase reactions. 
Continuing from Eq.~\eqref{eq:dRdt_active} along the lines of Lifshitz-Slyozov theory, we reproduce the same exact scaling but with an enhanced diffusivity,\footnote{Note that due to the transient for $R\ll\ell$ and possibly altered nucleation rates~\cite{ZiethenPRL23,ZwickerARXIV25}, the intercept $\langle R(0)\rangle^3_\mathrm{a}$ might differ from that of the passive case.}
\begin{equation}
    \langle R(t)\rangle^3-\langle R(0)\rangle^3_\mathrm{a}=\frac49D_\mathrm{eff}\lambda_{\s\g} t.\label{eq:active_ripening}
\end{equation}
This is our main result\,---\,activity enables Ostwald ripening to occur $(1+k_{\s\to\ns}/k_{\ns\to\s})$-fold faster compared to the passive case, Eq.~\eqref{eq:passive_ripening}, if the introduced activity is such that dense-phase reactions are slow enough. Interestingly, the faster the dilute-phase switching from sticky to nonsticky compared to the reverse process, the faster is the ripening rate. 

In addition to the restrictions on $s_\l$, there are two more limitations to this acceleration. First, the increase of $D_\mathrm{eff}$ occurs at the expense of having a lower sticky supersaturation, Eq.~\eqref{eq:supersatur}. Therefore, the saturation $\varphi^\infty$ must be sufficiently higher than $\phi_{\s\g}^\circ$ for the active solution to remain effectively supersaturated once the regime $R\gg\ell_\l,\ell_\g$ is reached, \textit{i.e.}, $\varphi^\infty_\mathrm{eff}-\phi_{\s\g}^\circ>0$. Second, biological systems are finite sized; \textit{e.g.}, a typical eukaryotic cell is $\sim10\mathrm{\mu m}$. For reasonable protein diffusivities $D_\g\sim10^{-1}-10^2\mathrm{\mu m}^2\mathrm{sec}^{-1}$ and conversion rates $k_{i\to j}\sim10^{-3}-10^{-1}\mathrm{sec}^{-1}$~\cite{MartinezCalvoARXIV24}, the distance between reactions varies between $\ell_\l,\ell_\g\sim10^{-1}-10^2\mathrm{\mu m}$. Therefore, the system size must be sufficient to attain the $R\gg\ell_\l,\ell_\g$ regime. 

It may not trivial to realize the above set of assumptions. Nonetheless, we argue for the feasibility of accelerating ripening by presenting recent experimental results as evidence for a workable protocol~\cite{TenaSolsonaCSC21}. There, a strongly hydrophobic (sticky) molecule (so $\phi_{\s\g}^\circ\to0$) converts to a hydrophilic (nonsticky) form upon exposure to water. Contrarily, a specialized fuel molecule converts the hydrophilic species through a sequence of reactions back into the hydrophobic from. Since the sticky species creates a hydrophobic condensate impenetrable to water, it is not lost by conversion at all inside droplets ($s_\l=0$). At the same time, in the dilute phase, the loss rate, $k_{\s\to\ns}$, is faster than the formation rate, $k_{\ns\to\s}$, so an acceleration factor of up to a $100$-fold is reported~\cite{TenaSolsonaCSC21}. 

\begin{figure}[t!]
    \centering
    \includegraphics[width=0.99\linewidth]{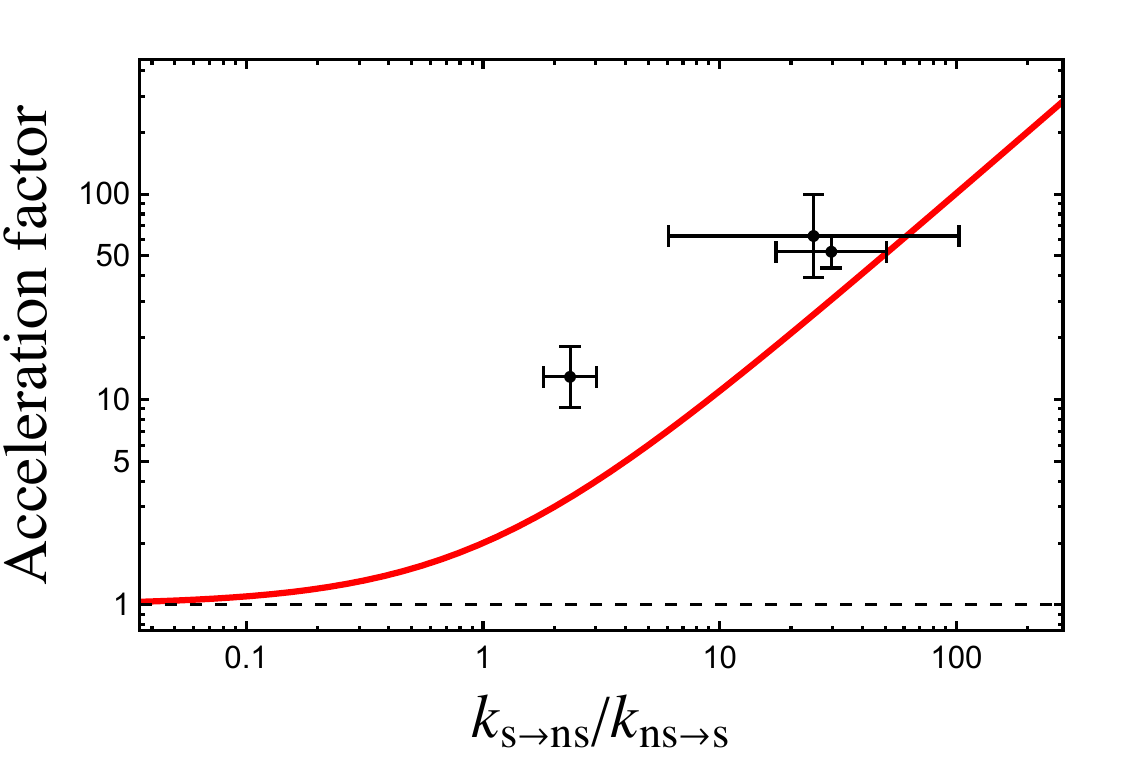}
    \caption{Active acceleration of Ostwald ripening. The solid red curve is the theoretical prediction of the acceleration factor\,---\,the ratio between the active ripening rate, Eq.~\eqref{eq:active_ripening}, and the passive one, Eq.~\eqref{eq:passive_ripening}, given by $D_\mathrm{eff}/D_\g=1+k_{\s\to\ns}/k_{\ns\to\s}$. The experimental data points~\cite{TenaSolsonaCSC21} correspond to a case where the fuel concentration was kept approximately constant. No fitting parameters were employed.}
    \label{fig:acceleration}
\end{figure}

In Fig.~\ref{fig:acceleration} we plot our theoretical prediction along with the experimental results~\cite{TenaSolsonaCSC21}, with no fitting parameters.\footnote{These three datapoints are obtained from experiments in which the fuel concentration was kept constant. Seven more datapoints are available from experiments where all of the fuel was introduced at the beginning,~\cite{TenaSolsonaCSC21} so during ripening only degradation of the hydrophobic species takes place. In that case, the accelerated ripening is enabled by accelerated shrinkage of small droplets. Our theory is quasistatic, and hence does not apply to this regime.} The acceleration factor is the ratio between slopes of the growth law of $R^3$ in the active and passive cases. Since the hydrophobic species does not exist without the fuel molecule, the growth rates for the passive cases are estimated via the Lifshitz-Slyozov law using measurements of surface tensions and solubilities. The agreement is not perfect, likely due to the approximations involved in estimating the passive growth rate, as well as $k_{\ns\to\s}$, and the protocol of periodic addition of the fuel that drives $k_{\ns\to\s}$. At the same time, the overall acceleration trend is reassuringly reproduced. Thus, we are provided with evidence that this singular means of accelerating ripening is indeed experimentally realizable~\cite{TenaSolsonaCSC21}.

To summarize, in this work, we theoretically studied the possibility of accelerating Ostwald ripening via chemical activity. We employed standard tools from reaction-diffusion and Lishitz-Slyozov theories to obtain explicit expressions for the diffusion-limited growth of active droplets under ripening. Indeed, if growth-hindering sticky-to-nonsticky reactions inside droplets can be avoided, we find acceleration of Ostwald ripening by an arbitrarily large factor, but with the same $R\sim t^{1/3}$ scaling imposed by particle conservation. Owing to the generality of our theory, we posit that this scenario in which reactions only occur in the dilute phase is indeed the only means to achieve accelerated Ostwald ripening. Our theoretical claims are supported by recent experiments in an organochemical solution~\cite{TenaSolsonaCSC21}.

Our theory has ignored droplet diffusion and coalescence\,---\,another means with which droplet growth can occur. For instance, biomolecular condensates typically have a low surface tension, so growth by coalescence might in fact be the dominant effect driving their growth in cells~\cite{LeeNP2023}. To complicate matters further, growth by coalescense also exhibits $R\sim t^{1/3}$ scaling~\cite{SiggiaPRA79,CatesBOOK17}, so distinguishing among growth by ripening from coalescence typically requires dynamic microscopy. Clearly, the interplay of reactions with droplet coalescence is a relevant future direction, as is the competition between the two growth mechanisms in active systems. We also ignored many other cellular complexities which interact with condensates~\cite{ZwickerARXIV25}. In the context of the pyrenoid, these include membrane wetting mediated by molecular tethers~\cite{GrandPreARXIV24} and the fact that two interacting components are required for droplet formation (Rubisco and EPYC1), rather than just a single ``sticky'' particle~\cite{ShanPLANT23}.

On a theoretical level, our work is a manifestation of an overall theme in nonequilibrium statistical mechanics concerning the ability to accelerate thermodynamic relaxation processes. One extensively studied example is the Mpemba effect, where an out-of-equilibrium system's initial condition is chosen so it will undergo relaxation along a faster trajectory~\cite{LuPNAS17}. Other accelerated phenomena include shortcuts to adiabaticity~\cite{GueryOdelinRMP19,TorronteguiPRA17,HoltzmanPRL25}, escape by active particles~\cite{WoillezJSTAT20,WoillezPRL19}, and search processes under stochastic resetting~\cite{EvansJPA20,EvansJPA11,PalPRL17}. In this context, it is important to stress that not all types or degrees of acceleration are physically achievable. Although activity typically breaks time-reversal symmetry (as in our case), any accelerated process is still constrained by remaining unbroken conservation laws.

On a practical level, to our knowledge, the acceleration theory we report has only been achieved in a chemical system~\cite{TenaSolsonaCSC21}. Nevertheless, we posit that realizing accelerated Ostwald ripening in biocondensates is plausible. For example, advances in synthetic biology allow the design of a phosphorylation-dependent switchable protein (like EPYC1) and a cognate kinase and phosphatase that are excluded from droplets~\cite{DaiNRB23}, thus preventing conversion reactions in the dense phase. If rapid coarsening can be turned on or off at will, \textit{e.g.}, by optogenetics~\cite{BrumbaughNC24}, the resulting size control of condensates would enable novel approaches to metabolic control~\cite{BananiNRCB17,ShinSCI17}.

To conclude, it is well established that living systems consume energy to carry out biological processes~\cite{MiloBOOK2015}. By driving molecular machinery ``uphill'' against chemical potential differences and thus breaking detailed balance, these active processes oppose relaxation to thermodynamic equilibrium. On the other hand, when relaxation to equilibrium is desired, it may be beneficial to accelerate the process. This fact is showcased by the ubiquity of enzymatic catalysis\,---\,a universal means to accelerate biological reactions~\cite{BergBOOK23}. Accelerating processes such as phase separation studied here but also membrane fission~\cite{ZhangJCP17,SpencerNC24} and fusion~\cite{McNewJCB00,KozlovskyBPJ02,ShendrikFD25}, DNA hybridization~\cite{OuldridgeNAR13}, and ligand-receptor binding~\cite{PangARB17} may be advantageous as well. It remains an exciting yet understudied question when and where cells harness energy to accelerate thermodynamic relaxation.

\section*{Acknowledgements}
We thank Job Boekhoven and Alejandro Mart\'inez-Calvo for illuminating discussions. B.S. acknowledges support from the Princeton Center for Theoretical Science. This project has been made possible in part by the Chan Zuckerberg Initiative DAF\,---\,an advised fund of Silicon Valley Community Foundation\,---\,grant number DAF2024-342781, NIH grant number R01GM140032, and NSF grant number MCB-2410354.

%


\renewcommand{\theequation}{S\arabic{equation}}
\setcounter{equation}{0}
\renewcommand{\thefigure}{S\arabic{figure}}
\setcounter{figure}{0}
\renewcommand{\thetable}{S\arabic{table}}
\setcounter{table}{0}
\renewcommand{\thesection}{S\Roman{section}}
\setcounter{section}{0}


\renewcommand{\o}{\mathrm{o}}
\newcommand{\T}{\mathrm{T}}
\newcommand{\bbar}[1]{\bar{\bar{#1}}}
\newcommand{\vecr}{\mathbf{r}}
\newcommand{\grad}{\boldsymbol{\nabla}}

\begin{widetext}
\newpage

\section*{Supplementary Material: Accelerated Ostwald ripening by chemical activity}
\section{Thermodynamic prerequisites and the Flory-Huggins free energy}\label{sec:thermo}

Most of the generic conclusions drawn in the main text for the equilibrium properties of phase-separating systems are motivated by the Flory-Huggins (FH) model~\cite{HugginsJCP41,FloryJCP42}. In this section, we first present the general thermodynamic principles underlying phase separation of a binary fluid (sticky solute and inert solvent). We proceed to connect the surface tension to the mesoscopic field theory considered in the main text. We then show concrete results for the FH free energy to provide intuition and to further support the assumptions concerning the volume fractions of each phase. We conclude by extending all these results to the case of a ternary fluid (sticky, nonsticky, and inert solvent) which is the case studied in the main text. 

\subsection{Thermodynamics of coexistence}

First, we outline the general thermodynamic principles governing phase-separating systems. We consider a material possessing a free energy $F$, which is a function of temperature $T$, volume $V$, and number of particles $\{N_i\}$, where $i=\s$ denotes the sticky particles and $i=\o$ the remaining inert and indistinguishable biomolecules of the cell. (We will add in the nonsticky species, $i=\ns$, in Sec.~\ref{sec:nonsticky} below). We suppose that the total local concentration of biomolecules (mainly proteins and RNAs) in the cytoplasm is fixed by strong osmotic constraints, hence, $\varrho=(N_\s+N_\o)/V=\mathrm{const}$. This imposes an effective incompressibility condition on the dissolved molecules,
\begin{equation}
    \phi_\s+\phi_\o=1,\label{eq:incomp2}
\end{equation}
where $\phi_i=N_i/(V\varrho)$ is the volume fraction of the $i$th species. For brevity, we refer to $\phi_i$ as a unitless concentration in the main text. Thus, $\phi_\o$ is uniquely determined by $\phi_\s$, as well as $N_\o$ by $N_\s$ and $V$. Furthermore, we assume that the temperature $T$ is also held constant. Upon phase separation, only $V$ and $N_\s$ vary, and so we will suppress the dependence of the free energy $F$ on the constants $T$ and $\varrho$.

With the above setup, Euler's homogeneous functions principle allows us to write the free energy as~\cite{DoiBOOK13}
\begin{equation}
    F(V,N_\s)=V\varrho \kt f\left(\frac{N_\s}{V\varrho}\right),\label{eq:free_energy}
\end{equation}
where $f(\phi_\s)$ is the total normalized free-energy per molecule (including both sticky and inert species). The sticky chemical potential and pressure are given by
\begin{subequations}
\begin{eqnarray}
    \frac{\mu_\s(\phi_\s)}\kt
    &=&\frac{1}\kt\left(\frac{\partial F}{\partial N_\s}\right)_{V}
    =f'(\phi_\s).\\
    \frac{P(\phi_\s)}{\varrho \kt}
    &=&-\frac{1}{\varrho\kt}\left(\frac{\partial F}{\partial V}\right)_{N_\s}
    =\phi_\s f'(\phi_\s)-f(\phi_\s),\label{eq:pressure}
\end{eqnarray}
\end{subequations}
where $f'\equiv\partial f/\partial\phi_\s$ for shorthand.

Phase separation is possible whenever the free energy has more than one stable volume fraction. We denote the coexistence volume fraction of the dense phase (sticky-rich, ``liquid'') by $\phi_{\s\l}^\circ$ and of the dilute phase (sticky-poor, ``gas'') by $\phi_{\s\g}^\circ$. When two phases in contact are in equilibrium, the sticky chemical potential and the pressure of the two phases equate~\cite{DoiBOOK13},
\begin{subequations}
\label{eq:coex0}
\begin{eqnarray}
    \mu_\s(\phi_{\s\l}^\circ)&=&\mu_\s(\phi_{\s\g}^\circ),\\
    P(\phi_{\s\l}^\circ)&=&P(\phi_{\s\g}^\circ).
\end{eqnarray}
\end{subequations}
These two equations determine the volume fractions of each phase. Furthermore, a known total amount of sticky material $N_\s$ allows one to determine the volume fraction of the dense phase $\zeta_\l=1-\zeta_\g$ when phase separation is complete, $\zeta_\l\phi_{\s\l}^\circ+(1-\zeta_\l)\phi_{\s\g}^\circ=N_\s/(V\varrho)$.

To treat a droplet, the liquid phase is assumed to be finite sized and spherical with radius $R$. As a result, we must account for the surface tension $\sigma$, which leads to a curvature-induced Laplace pressure, $2\sigma/R$~\cite{deGennesBOOK03}. Thus, The coexistence volume fractions modify to $\phi_{\s\l}^\L$ and $\phi_{\s\g}^\L$, satisfying
\begin{subequations}
\label{eq:coexL}
\begin{eqnarray}
    \mu_\s(\phi_{\s\l}^\L)&=&\mu_\s(\phi_{\s\g}^\L),\\
    P(\phi_{\s\l}^\L)&=&P(\phi_{\s\g}^\L)+\frac{2\sigma}{R}.
\end{eqnarray}
\end{subequations}
Assuming the droplet is large (so Laplace pressure is small), we approximate $\phi_{\s\alpha}^\L\simeq\phi_{\s\alpha}^\circ+\lambda_{\s\alpha}/R$ in each phase $\alpha=\l,\g$ (where $\{\lambda_{\s\alpha}\}$ are Laplace-pressure correction constants), and expand Eqs.~\eqref{eq:coexL} for small $\lambda_{\s\alpha}/R$. This yields Eqs.~\eqref{eq:coex0} for the leading order, and the following next-order equations: 
\begin{subequations}
\label{eq:coexApprx}
\begin{eqnarray}
    f''(\phi_{\s\l}^\circ)\lambda_{\s\l}&=&f''(\phi_{\s\g}^\circ)\lambda_{\s\g},\\
    \phi_{\s\l}^\circ f''(\phi_{\s\l}^\circ)\lambda_{\s\l}&=&\phi_{\s\g}^\circ f''(\phi_{\s\g}^\circ)\lambda_{\s\g}+\frac{2\sigma}{\varrho\kt}.
\end{eqnarray}
\end{subequations}
The solution to these equations is 
\begin{equation}
    \lambda_{\s \alpha}=\frac{1}{\phi_{\s\l}^\circ-\phi_{\s\g}^\circ}\frac{2\sigma}{\varrho\kt f''(\phi_{\s\alpha}^\circ)}=\frac{2(\phi_{\s\alpha}^\circ)^2\beta(\phi_{\s\alpha}^\circ)\sigma}{\phi_{\s\l}^\circ-\phi_{\s\g}^\circ}.\label{eq:Laplace_lambda}
\end{equation}
where, using Eq.~\eqref{eq:pressure}, we identified the isothermal compressibility coefficient~\cite{AtkinsBOOK23},
\begin{equation}
    \beta(\phi_\s)\equiv-\frac1V\left(\frac{\partial V}{\partial P}\right)_{N_\s}=\frac1{\varrho\kt\phi_\s^2f''(\phi_\s)}.    
\end{equation}
Indeed, the higher the surface tension $\sigma$ and compressibility $\beta$ (the susceptibility of each phase to compression due to the extra Laplace pressure), the larger is the increase in sticky volume fraction due to Laplace pressure.

For completeness, we conclude by presenting the following common simplification of Eq.~\eqref{eq:Laplace_lambda}: In a typical system, the liquid phase is much denser than the gas phase, $\phi_{\s\l}^\circ-\phi_{\s\g}^\circ\simeq \phi_{\s\l}^\circ$, and furthermore the latter can be approximated as ideal, $\beta(\phi_{\s\g}^\circ)\simeq1/(\varrho\kt\phi_{\s\g}^\circ)$. Thus, the gas-phase volume fractions follow an Ostwald-Freundlich-type equation,
\begin{equation}
    \frac{\phi_{\s\g}^\L}{\phi_{\s\g}^\circ}\simeq1+\frac{1}{\varrho\kt\phi_{\s\l}^\circ}\frac{2\sigma}R\simeq\exp\left[\frac{2v_\mathrm{m}\sigma}{\kt R}\right],
\end{equation}
where $v_\mathrm{m}=(\varrho\phi_{\s\l}^\circ)^{-1}$ is the volume per sticky particle in the liquid phase. 

\subsection{Surface tension}\label{sec:surface_tension}

In our coarse-grained view of a phase-separating system, a given particle configuration is modeled via a volume-fraction field. To leading order in small-wavelength excitations, we promote the homogeneous free energy (Eq.~\eqref{eq:free_energy}) into a free energy functional~\cite{CahnJCP58,MaoSM19},
\begin{equation}
    F[\phi_\s(\vecr)]=\varrho \kt \int\d\vecr\left[f(\phi_\s(\vecr))+\frac{\kappa_\s}2|\grad\phi_\s(\vecr)|^2\right],\label{eq:free_energy_functional}
\end{equation}
where we have included the leading-order energetic penalty in volume-fraction gradients, whose magnitude is controlled by $\kappa_\s$. Equation~\eqref{eq:free_energy_functional} coincides with Eq.~\eqref{eq:free_energy} for a homogeneous medium. Incompressibility (Eq.~\eqref{eq:incomp2}) is assumed at every position, $\phi_\s(\vecr)+\phi_\o(\vecr)=1$. In this section, we show how the macroscopic surface tension $\sigma$ can be related to the microscopic penalty $\kappa_\s$~\cite{CahnJCP58,MaoSM19}.

We compute the surface tension due to a flat interface between two semi-infinite bulks. (Any curvature might, in principle, result in an additional $\mathcal{O}(R^{-1})$ contribution to the surface tension, which is already subleading when inserted into the Laplace-pressure corrections to the volume fractions.) We denote the axis perpendicular to the interface as $x$, with the interface lying on $x=0$, and we arbitrarily designate the left side as the gas, $\phi_\s(-\infty)=\phi_{\s\g}^\circ$, and right side to be the liquid $\phi_\s(\infty)=\phi_{\s\l}^\circ$.

In equilibrium, the local chemical potential, $\delta F/\delta[\varrho\phi_\s(x)]$ (where $\varrho\phi_\s$ is the local particle density), is uniform everywhere so no particle transport would occur,
\begin{equation}
   f'(\phi_\s(x))-\kappa_\s\frac{\d^2\phi_\s(x)}{\d x^2}=\frac{\mu_\s(\phi_{\s\g}^\circ)}\kt,\label{eq:uniform_chem_pot}
\end{equation}
where $\mu_\s(\phi_{\s\g}^\circ)=\mu_\s(\phi_{\s\l}^\circ)$ is the chemical potential deep in each bulk. We multiply each side of the equation by $\d\phi_\s(x)/\d x$, and integrate from $-\infty$ to $x$,
\begin{equation}
    f(\phi_\s(x))-\frac{\kappa_\s}{2}\left[\frac{\d\phi_\s(x)}{\d x}\right]^{2}-f(\phi_{\s\g}^{\circ})=\frac{\mu_\s(\phi_{\s\g}^{\circ})}\kt[\phi_\s(x)-\phi_{\s\g}^{\circ}].\label{eq:uniform_chem_pot_mod}
\end{equation}

The surface tension is the accumulated excess pressure due to the interface~\cite{CahnJCP58,MaoSM19}\,---\,\textit{i.e.}, the difference between the local pressure, $\varrho\phi_\s(x)\mu_\s(\phi_{\s\g}^\circ)-\varrho\kt\{f(\phi_\s(x))+(1/2)\kappa_\s[\d\phi_\s(x)/\d x]^2\}$ (where $\mu_\s(\phi_{\s\g}^\circ)$ is the uniform chemical potential, $\varrho\phi_\s(x)$ is the local density, and the second term is the local free energy density, with the three inserted in Eq.~\eqref{eq:pressure}), relative to the pressure without the interface, $\varrho\phi_{\s\g}^\circ \mu_\s(\phi_{\s\g}^{\circ})-\varrho\kt f(\phi_{\s\g}^\circ)=\varrho\phi_{\s\l}^\circ \mu_\s(\phi_{\s\l}^{\circ})-\varrho\kt f(\phi_{\s\l}^\circ)$; \textit{cf.} Eq.~\eqref{eq:pressure}. Explicitly,
\begin{equation}
    \sigma=-\varrho \kt \int_{-\infty}^\infty\d x\Biggl(\left[\phi_\s(x)\frac{\mu_\s(\phi_{\s\g}^\circ)}\kt-\left\{f(\phi_\s(x))+\frac{\kappa_\s}{2}\left[\frac{\d\phi_\s(x)}{\d x}\right]^{2}\right\}\right]-\left[\phi_{\s\g}^\circ \frac{\mu_\s(\phi_{\s\g}^\circ)}\kt-f(\phi_{\s\g}^\circ)\right]\Biggr).\label{eq:surface_tension_def}
\end{equation}
Expressing $(\kappa_\s/2)[\d\phi_\s(x)/\d x]^2$ via Eq.~\eqref{eq:uniform_chem_pot_mod}, we simplify Eq.~\eqref{eq:surface_tension_def} as
\begin{eqnarray}
    \frac\sigma{\varrho\kt}&=&\sqrt{2\kappa_\s}\int_{-\infty}^\infty\d x\frac{\d\phi_\s(x)}{\d x}\left\{[f(\phi_\s(x))-f(\phi_{\s\g}^\circ)]-\frac{\mu_\s(\phi_{\s\g}^\circ)}\kt[\phi_\s(x)-\phi_{\s\g}^\circ]\right\}^{1/2}\nonumber\\
    &=&\sqrt{2\kappa_\s}\int_{\phi_{\s\g}^\circ}^{\phi_{\s\l}^\circ}\d a\left\{[f(a)-f(\phi_{\s\g}^\circ)]-\frac{\mu_\s(\phi_{\s\g}^\circ)}\kt(a-\phi_{\s\g}^\circ)\right\}^{1/2},\label{eq:surface_tension_res}
\end{eqnarray}
where we changed variables of integration from $x\in(-\infty,\infty)$ to $a=\phi(x)\in(\phi_{\s\g}^\circ,\phi_{\s\l}^\circ)$. Thus we find the scaling $\sigma\sim\sqrt{\kappa_\s}$, which could have also been found from scaling arguments. The integral brings forth the dependence on the underlying energetics of the bulks.

\subsection{Flory-Huggins free energy}\label{sec:FHmodel}

As a motivating example to support our choices of volume fractions and the general intuition in the main text, we specialize to the FH free energy~\cite{HugginsJCP41,FloryJCP42}, which is widely used in modeling phase separation of simple liquids. It has two contributions; one coming from the entropy of mixing, and the other from sticky-sticky interactions,
\begin{equation}
    f(\phi_\s)=\phi_\s\ln\phi_\s+(1-\phi_\s)\ln(1-\phi_\s)+\chi\phi_\s(1-\phi_\s),\label{eq:FHfree_energy}
\end{equation}
where $\kt(-\chi)$ accounts for the attractive ($\chi>0$) energy between sticky particles. We assumed for simplicity that the molecular volumes of the species~\cite{ZwickerARXIV25}, as well as their polymeric lengths~\cite{deSouzaJCP24}, are equal. The chemical potential and pressure are given by
\begin{subequations}
\begin{eqnarray}
    \frac{\mu_\s(\phi_\s)}\kt&=&\ln\phi_\s-\ln(1-\phi_\s)+\chi(1-2\phi_\s),\\
    \frac{P(\phi_\s)}{\varrho \kt}&=&-\ln(1-\phi_\s)-\chi\phi_\s^2.
\end{eqnarray}
\end{subequations}

We find the volume fractions of each phase in coexistence. Phase separation occurs in this model for $\chi>2$~\cite{deSouzaJCP24}. To model strong phase separation, $(1-\phi_{\s\l}^\circ),\phi_{\s\g}^\circ\ll1$, we consider $\chi\gg1$. Upon solving Eqs.~\eqref{eq:coex0} for the FH free energy under the ansatz $\phi_{\s\l}^\circ\simeq1$ and $\phi_{\s\g}^\circ\simeq0$, we find
\begin{subequations}
\label{eq:FHcoex_volume_fraction}
\begin{eqnarray}
    \phi_{\s\l}^\circ&=&1-e^{-\chi},\\
    \phi_{\s\g}^\circ&=&e^{-\chi},
\end{eqnarray}
\end{subequations}
where corrections are of order $\chi e^{-2\chi}$, which are thus consistent with $\chi\gg1$. Inserting these volume fractions in Eq.~\eqref{eq:Laplace_lambda}, we find
\begin{equation}
    \lambda_{\s\l}=\lambda_{\s\g}\equiv\lambda=\frac{2\sigma e^{-\chi}}{\varrho\kt},\label{eq:FHlambda}
\end{equation}
with, again, corrections of order $\chi e^{-2\chi}$. 

Lastly, we compute the surface tension $\sigma$ for FH in the large-$\chi$ limit. First, since $\kappa_\s$ is an energy penalty due to gradients, we expect $\kappa_\s=\chi\xi^2$, where $\xi$ is the effective lengthscale over which the microscopic sticky-sticky short-range potential is acting (whereas $\chi$ controls that potential's magnitude). Thus, using Eq.~\eqref{eq:surface_tension_def}, we find for $\chi\gg1$
\begin{equation}
    \frac\sigma{\varrho\kt}=\frac\pi{4\sqrt2}\xi\left[\chi-4\ln\left(\frac4e\right)\right],\label{eq:FH_surface_tension}
\end{equation}
where higher-order corrections are $\chi^2e^{-2\chi}$. Overall, 
\begin{equation}
    \lambda=\frac\pi{2\sqrt2}\xi e^{-\chi}\left[\chi-4\ln\left(\frac4e\right)\right].
\end{equation}
We see, therefore, that the large droplet approximation is consistent when $R\gg\chi\xi$, which is reasonable as droplets should be much larger than molecular lengthscales. Moreover, the scaling $\lambda\sim\xi\chi e^{-\chi}$ indicates that the surface tension increases more slowly with $\chi$ than the compressibility of the liquid phase decays. Thus, perhaps counterintuitively, Ostwald ripening occurs slower for systems which energetically tend more strongly to phase separate.

\subsection{Ternary mixtures}\label{sec:nonsticky}

We now extend all formulas presented above to the case where a third, nonsticky component is present. The simplest conceptual way to think of the (inert) nonsticky species is as if one has labeled a given fraction of the inert proteome~\cite{KampenBOOK84}. Thus, the total available protein concentration is still fixed at $\varrho=(N_\s+N_\ns+N_\o)/V$, and the incompressibility condition (Eq.~\eqref{eq:incomp2}) is modified to include the nonsticky volume fraction $\phi_\ns$,
\begin{equation}
    \phi_\s+\phi_\ns+\phi_\o=1.\label{eq:incomp3}
\end{equation}
The free energy (Eq.~\eqref{eq:free_energy}), therefore, is rewritten as
\begin{equation}
    F(V,N_\s,N_\s)=V\varrho \kt f\left(\frac{N_\s}{V\varrho},\frac{N_\ns}{V\varrho}\right),
\end{equation}
where $f(\phi_\s,\phi_\ns)$ is the amended normalized free-energy per molecule. The two chemical potentials ($i=\s,\ns$) and pressure are now given by
\begin{subequations}
\begin{eqnarray}
    \frac{\mu_i(\phi_\s,\phi_\ns)}\kt
    &=&\frac{\partial f}{\partial\phi_i}\\
    \frac{P(\phi_\s,\phi_\ns)}{\varrho \kt}
    &=&\phi_\s \frac{\partial f}{\partial\phi_\s}+\phi_\ns \frac{\partial f}{\partial\phi_\ns}-f(\phi_\s,\phi_\ns).
\end{eqnarray}
\end{subequations}

When the system phase separates, per the Gibbs phase rule~\cite{AtkinsBOOK23}, only three thermodynamic constraints are imposed on the four coexistence volume fractions $\{\phi_{i\alpha}^\L=\phi_{i\alpha}^\circ+\lambda_{i\alpha}/R\}$ (where $i=\s,\ns$ and $\alpha=\l,\g$),
\begin{subequations}
\label{eq:coexL3}
\begin{eqnarray}
    \mu_\s(\phi_{\s\l}^\L,\phi_{\ns\l}^\L)&=&\mu_\s(\phi_{\s\g}^\L,\phi_{\ns\g}^\L),\\
    \mu_\ns(\phi_{\s\l}^\L,\phi_{\ns\l}^\L)&=&\mu_\ns(\phi_{\s\g}^\L,\phi_{\ns\g}^\L),\label{eq:coexL3ns}\\
    P(\phi_{\s\l}^\L,\phi_{\ns\l}^\L)&=&P(\phi_{\s\g}^\L,\phi_{\ns\g}^\L)+\frac{2\sigma}R.
\end{eqnarray}
\end{subequations}
All four volume fractions (as well as the four Laplace-pressure corrections) can be deduced by imposing material conservation across both phases when phase separation is complete, $\zeta_\l\phi_{\s\l}^\circ+(1-\zeta_\l)\phi_{\s\g}^\circ=N_\s/(V\varrho)$ and $\zeta_\l\phi_{\ns\l}^\circ+(1-\zeta_\l)\phi_{\ns\g}^\circ=N_\ns/(V\varrho)$, where $\zeta_\l=1-\zeta_\g$ is the volume fraction of the dense phase (total five equations with five unknowns $\{\phi_{i\alpha}^\L\}$ and $\zeta_\l$), or by flux matching during droplet growth (Eq.~(3) of the main text, for a total of five equations with five unknowns $\{\phi_{i\alpha}^\L\}$ and $\d R/\d t$).

We note that, since it is inert, the nonsticky species should not have a volume-fraction gradient penalty in the free energy functional. Thus, the free energy functional (Eq.~\eqref{eq:free_energy_functional}) becomes
\begin{equation}
    F[\phi_\s(\vecr),\phi_\ns(\vecr)]=\varrho \kt \int\d\vecr\left[f(\phi_\s(\vecr),\phi_\ns(\vecr))+\frac{\kappa_\s}2|\grad\phi_\s(\vecr)|^2\right].
\end{equation}
When estimating the surface tension at the interface between two semi-infinite bulks, the uniformity of both chemical potentials (\textit{cf.} Eq.~\eqref{eq:uniform_chem_pot}) must be required,
\begin{subequations}
\label{eq:uniform_chem_pot2}
\begin{eqnarray}
    \frac{\partial f}{\partial\phi_\s}\Biggr|_{\phi_\s(x),\phi_\ns(x)}-\kappa_\s\frac{\d^2\phi_\s(x)}{\d x^2}&=&\frac{\mu_\s(\phi_{\s\g}^\circ,\phi_{\ns\g}^\circ)}\kt,\\
    \frac{\partial f}{\partial\phi_\ns}\Biggr|_{\phi_\s(x),\phi_\ns(x)}&=&\frac{\mu_\ns(\phi_{\s\g}^\circ,\phi_{\ns\g}^\circ)}\kt.\label{eq:uniform_chem_pot_ns}
\end{eqnarray}
\end{subequations}
Note that Eq.~\eqref{eq:uniform_chem_pot_ns} implies that $\phi_\ns(x)$ spatially rearranges in accord with the local $\phi_\s(x)$, so the positional dependence of the former is strictly acquired from the latter, $\phi_\ns(x)\to\phi_\ns(\phi_\s(x))$. By multiplying each of Eqs.~\eqref{eq:uniform_chem_pot2} by $\d\phi_i(x)/\d x$, summing the two equations, and integrating the result from $-\infty$ to $x$, we obtain
\begin{equation}
    f(\phi_\s(x),\phi_\ns(\phi_\s(x)))-\frac{\kappa_\s}{2}\left[\frac{\d\phi_\s(x)}{\d x}\right]^{2}-f(\phi_{\s\g}^{\circ},\phi_{\ns\g}^{\circ})=\frac{\mu_\s(\phi_{\s\g}^{\circ},\phi_{\ns\g}^{\circ})}\kt[\phi_\s(x)-\phi_{\s\g}^{\circ}]-\frac{\mu_\ns(\phi_{\s\g}^{\circ},\phi_{\ns\g}^{\circ})}\kt[\phi_\ns(\phi_\s(x))-\phi_{\ns\g}^{\circ}].
\end{equation}
Via a similar procedure to Sec.~\ref{sec:surface_tension}, the surface tension can be found as
\begin{equation}
    \frac\sigma{\varrho\kt}=\sqrt{2\kappa_\s}\int_{\phi_{\s\g}^\circ}^{\phi_{\s\l}^\circ}\d a\Biggl\{[f(a,\phi_\ns(a))-f(\phi_{\s\g}^\circ,\phi_{\ns\g}^\circ)]-\frac{\mu_\s(\phi_{\s\g}^\circ,\phi_{\ns\g}^\circ)}\kt(a-\phi_{\s\g}^\circ)-\frac{\mu_\ns(\phi_{\s\g}^\circ,\phi_{\ns\g}^\circ)}\kt[\phi_\ns(a)-\phi_{\ns\g}^\circ]\Biggr\}^{1/2}.\label{eq:surface_tension_res2}
\end{equation}

To make the case with nonsticky particles concrete, we amend the FH free energy (Eq.~\eqref{eq:FHfree_energy}) as
\begin{equation}
    f(\phi_\s,\phi_\ns)=\phi_\s\ln\phi_\s+\phi_\ns\ln\phi_\ns+(1-\phi_\s-\phi_\ns)\ln(1-\phi_\s-\phi_\ns)+\chi\phi_\s(1-\phi_\s),\label{eq:FHfree_energy_3}
\end{equation}
where the ``labeling'' of nonsticky particles results in an additional entropy contribution~\cite{KampenBOOK84}. The internal energy term has remained unchanged as only the sticky particles are noninert. Once again, all species have been assumed to have the same molecular volumes~\cite{ZwickerARXIV25} and number of monomers in the case of polymers~\cite{deSouzaJCP24}. The chemical potentials and pressure are given by
\begin{subequations}
\begin{eqnarray}
    \frac{\mu_\s(\phi_\s,\phi_\ns)}\kt&=&\ln\phi_\s-\ln(1-\phi_\s-\phi_\ns)+\chi(1-2\phi_\s),\\
    \frac{\mu_\ns(\phi_\s,\phi_\ns)}\kt&=&\ln\phi_\ns-\ln(1-\phi_\s-\phi_\ns),\\
    \frac{P(\phi_\s,\phi_\ns)}{\varrho \kt}&=&-\ln(1-\phi_\s-\phi_\ns)-\chi\phi_\s^2.
\end{eqnarray}
\end{subequations}

While only three thermodynamic constraints are given at coexistence (Eqs.~\eqref{eq:coexL3}), we can make progress in the large-$\chi$ regime for the FH-model nonetheless. First, via Eq.~\eqref{eq:coexL3ns} we confirm that the nonsticky species remain equivalent to the inert solvent despite the labeling. Namely, the ratio between the nonsticky volume fraction ($\phi_\ns$) and the solvent one ($\phi_\o=1-\phi_\s-\phi_\ns$) remains fixed across both phases,
\begin{equation}
    \frac{\phi_{\ns\l}^\L}{1-\phi_{\s\l}^\L-\phi_{\ns\l}^\L}=\frac{\phi_{\ns\g}^\L}{1-\phi_{\s\g}^\L-\phi_{\ns\g}^\L}.\label{eq:equiv_ns_o}
\end{equation}
With this assurance, we proceed to posit, once again, that $\phi_{\s\l}^\circ\simeq1$ and $\phi_{\s\g}^\circ\simeq0$ for $\chi\gg1$, as the sticky protein is the one driving phase separation. Indeed, we once again find Eqs.~\eqref{eq:FHcoex_volume_fraction} and~\eqref{eq:FHlambda}, whereas the nonsticky volume fractions are found to satisfy $\phi_{\ns\l}^\L=\phi_{\ns\g}^\L(e^{-\chi}+\lambda/R)$, so
\begin{equation}
    \phi_{\ns\l}^\circ=\phi_{\ns\g}^\circ e^{-\chi}
\end{equation}
and
\begin{equation}
    \lambda_{\ns\l}=\lambda_{\ns\g}e^{-\chi}+\lambda\phi_{\ns\g}^\circ.
\end{equation}
Thus, per the Gibbs phase rule, only one unknown $\phi_{\ns\g}^\circ$(and its leading-order finite-droplet correction, $\lambda_{\ns\g}$) remain to be determined. For brevity, since we will find $\phi_{\ns\g}^\circ\ll1$, we set $\phi_{\ns\l}^\L\ll e^{-\chi}$ to be simply  $\phi_{\ns\l}^\L=0$ in the main text.

Lastly, we confirm that the FH surface tension (Eq.~\eqref{eq:surface_tension_res}) does not change due to ``labeling'' inert (nonsticky) species. First, Eq.~\eqref{eq:uniform_chem_pot_ns} simplifies for the FH free energy as $\phi_{\ns}(x)/[1-\phi_{\s}(x)-\phi_{\ns}(x)]=\phi_{\ns\g}^\circ/(1-\phi_{\s\g}^\circ-\phi_{\ns\g}^\circ)$,
which is the position-dependent analogue of the inert proteome/nonsticky protein equivalence argument, Eq.~\eqref{eq:equiv_ns_o}. Inserting it in Eq.~\eqref{eq:surface_tension_res2} with the FH free energy (Eq.~\eqref{eq:FHfree_energy_3}) results in Eq.~\eqref{eq:surface_tension_res} with the binary-fluid's FH free energy (Eq.~\eqref{eq:FHfree_energy}). Thus, indeed the surface tension is ignorant of inert-protein labels, so $\sigma$ is still given by Eq.~\eqref{eq:FH_surface_tension}.


\section{From volume-fraction field dynamics to droplet growth law}

In this section, we detail the full derivation of the droplet growth laws in the presence of reactions. Particularly, we list our approximations, the phenomena that should be avoided in the liquid phase to enable growth, derive the flux-matching equation (Eq.~(3) of the main text), extract $\phi_{\ns\g}^\L$ from it in the different limits of droplet size $R$, and from it the radius growth laws. Throughout, we reference the results of the FH free energy to support general intuition.

\subsection{Reaction-diffusion approach}

Here, we construct the reaction-diffusion equations that our analysis is based on. We denote with vectors and matrices
\begin{equation}
    \bar z=\left(\begin{array}{c}
         z_\s \\
         z_\ns 
    \end{array}\right),\qquad 
    \bbar Z=\left(\begin{array}{cc}
         Z_{\s\s} & Z_{\s\ns}\\
         Z_{\ns\s} & Z_{\ns\ns}
    \end{array}\right),
\end{equation}
to spare cumbersome notation indicating a summation over particle types. 

First, following linear-irreversible thermodynamics~\cite{deGrootBOOK84}, the volume-fraction fields $\bar\phi(\vecr,t)$ respond linearly to chemical potential gradients $\delta F/\delta[\varrho\bar\phi(\vecr,t)]$ via a (normalized) mobility matrix denoted $\varrho\kt \bbar\Gamma(\bar\phi(\vecr,t))$. Thus, the diffusive part of the dynamics is given by 
\begin{equation}
    \frac{\partial \bar\phi(\vecr,t)}{\partial t}=-\grad\cdot\bar{\mathbf{J}}(\vecr,t),\qquad \bar{\mathbf{J}}(\vecr,t)=\bbar\Gamma(\bar\phi(\vecr,t))\grad\left[\frac{\partial f}{\partial\bar\phi}\Biggr|_{\bar\phi(\vecr,t)}-\bbar\kappa\nabla^2\bar\phi(\vecr,t)\right],
\end{equation}
where $\kappa_{ij}=\delta_{i\s}\delta_{j\s}\kappa_\s$. 

Second, we suppose that a collection of $M_\mathrm{R}$ reactions take place, whose rates as a function of protein concentrations are $\{S_m(\bar\phi)\}_{m=1,\ldots,M_\mathrm{R}}$. Denoting the set of stoichiometric constants for each species in the $m$th reaction as $\bar\nu_m$, we find the reactions' contribution to the volume-fraction dynamics,
\begin{equation}
    \frac{\partial \bar\phi(\vecr,t)}{\partial t}=\sum_{m=1}^{M_\mathrm{R}}\bar\nu_mS_m(\bar\phi(\vecr,t)).\label{eq:react_prep}
\end{equation}
We stress that the reaction mechanism must conserve material. Namely, if $\s$ and $\ns$ react with some stoichiometric constants, they are bound to satisfy the same ratio among constants across all reactions. Otherwise, it would seem as if material is lost or gained spontaneously. If material is not conserved, this means that the two-component description is insufficient, and underlying ignored components would have to be considered. We assume that this is not the case, and the system is properly described as a two-component one. Hence, necessarily, we can rearrange Eq.~\eqref{eq:react_prep} as
\begin{equation}
    \sum_{m=1}^{M_{\mathrm{R}}}\bar\nu_mS_m(\bar\phi(\vecr,t))\equiv\bar\upsilon s(\bar{\phi}(\vecr,t)),
\end{equation}
where $s(\bar{\phi})$ is a single unified reaction rate corresponding to some choice of stoichiometric coefficients $\bar\upsilon$ satisfying the abovementioned material balance.

Upon combining both reactions and diffusion, we find
\begin{equation}
    \frac{\partial \bar\phi(\vecr,t)}{\partial t}=\grad\cdot\Biggl\{\bbar\Gamma(\bar\phi(\vecr,t))\grad\Biggl[\frac{\partial f}{\partial\bar\phi}\Biggr|_{\bar\phi(\vecr,t)}-\bbar\kappa\nabla^2\bar\phi(\vecr,t)\Biggr]\Biggr\}+\bar\upsilon s(\bar\phi(\vecr,t)).\label{eq:phi_dynamics}
\end{equation}
In the main text, we assume a one-to-one stoichiometric ratio between $\s$ and $\ns$ for the interconversion reactions. Here, we proceed with a general ratio. 
Since $\s$ is created from $\ns$ and vice versa, $\upsilon_\s$ and $\upsilon_\ns$ must be of opposite signs. Without loss of generality (up to rescaling of $s(\bar\phi)$), we choose $\upsilon_\s<0$ and $\upsilon_\ns=1$; $|\upsilon_\s|$ thus determines the stoichiometric ratio. (In the main text, $\upsilon_\s=-1$.) For later convenience, we also define $\bar\upsilon_\perp$ as the orthogonal vector to $\bar\upsilon$, so $\bar\upsilon_\perp^\T\bar\upsilon=0$. Again without loss of generality, we choose it as $\bar\upsilon_\perp=(1,|\upsilon_\s|)^\T$ (whereas $\bar\upsilon=(-|\upsilon_\s|,1)^\T$). 

Multiplying Eq.~\eqref{eq:phi_dynamics} by $\bar\upsilon_\perp^\T$, we identify a conserved field $\bar\upsilon_\perp^\T\bar\phi(\vecr,t)$, which is ignorant of the reactions. Next, we suppose that the nonsticky proteins are formed out of an initial sticky-only solution whose saturation was $\varphi^\infty$ (see main text). Thus, with the above choice for $\bar\upsilon_\perp$, the saturation is conserved via $\bar\upsilon_\perp^\T\bar\varphi^\infty=\varphi^\infty$. The two saturations $\bar\varphi^\infty=(\varphi^\infty_\s,\varphi^\infty_\ns)^\T$ far away from a droplet are determined from the reaction equilibrium, $s(\bar\varphi^\infty)=0$.

To make analytical progress from Eq.~\eqref{eq:phi_dynamics}, we introduce a number of standard simplifications~\cite{ZwickerARXIV25}, valid at a late stage of coarsening. First, the droplets are assumed to be sparse (the interdroplet distances are much greater than their radii), so we consider a single phase-separated droplet as if it is in isolation, diffusively exchanging material with the slightly-supersaturated dilute phase, $0<\phi_{\s\g}^\circ\lesssim\varphi^\infty_\s\ll1$. Assuming a large surface tension, the droplets are spherical at all times, with radius $R(t)$ much larger than the interface width $\xi$ (defined in Sec.~\ref{sec:FHmodel}). We split the volume fraction fields to droplet interior (liquid) and exterior (gas), $\bar\phi_\l(r,t)=\bar\phi(|\vecr-\vecr_\mathrm{CM}|<R(t),t)$ and $\bar\phi_\g(r,t)=\bar\phi(|\vecr-\vecr_\mathrm{CM}|>R(t),t)$, respectively, where $\vecr_\mathrm{CM}$ is the droplet-center coordinate. Assuming the reactions are slow enough not to interfere with surface thermodynamics~\cite{ChoJCP23,WeyerARXIV24}, the two volume fractions are always equilibrated at the interface, $\bar\phi_{\alpha}(R(t),t)=\bar\phi_\alpha^\L(R(t))$ ($\alpha=\l,\g$). Lastly, the droplets grow quasistatically, whereby the volume fraction fields $\{\bar\phi_\alpha(r;R)\}$ are found for a quasi-steady state of a fixed radius $R$, and then droplet growth rate is inferred from the resulting steady diffusive flux, $\bar J_\alpha(R)$. 

\subsection{Droplet growth in terms of particle flux}

Prior to continuing, we derive Eq.~(3) of the main text, relating the droplet-radius growth rate to the particle fluxes. We integrate Eq.~\eqref{eq:phi_dynamics} (assuming spherical symmetry) over the entire container,
\begin{equation}
    \frac\d{\d t}\int_0^\infty \d rr^2\bar\phi(r,t)=\bar\upsilon\int_0^\infty \d rr^2s(\bar\phi(r,t)).\label{eq:R_deriv_1}
\end{equation}
where the flux at the boundaries is zero in a closed system. Alternatively, we split the left-hand side into liquid and gas regions, and obtain using Eq.~\eqref{eq:phi_dynamics} and the Leibniz integral rule,
\begin{eqnarray}
    \frac\d{\d t}\int_0^\infty \d rr^2\bar\phi(r,t)&=&\frac\d{\d t}\int_0^{R^-(t)}\d rr^2\bar\phi(r,t)+\frac\d{\d t}\int_{R^+(t)}^\infty \d rr^2\bar\phi(r,t)\nonumber\\
    &=&\frac{\d R(t)}{\d t}\bar\phi(R^-(t),t)-\int_0^{R^-(t)}\d rr^2\frac1{r^2}\frac\partial{\partial r}\left[r^2\bar J(r,t)\right]+\bar\upsilon\int_0^{R^-(t)}\d rr^2s(\bar\phi(r,t))\nonumber\\
    &&-\frac{\d R(t)}{\d t}\bar\phi(R^+(t),t)-\int_{R^+(t)}^\infty\d rr^2\frac1{r^2}\frac\partial{\partial r}\left[r^2\bar J(r,t)\right]+\bar\upsilon\int_{R^+(t)}^\infty\d rr^2s(\bar\phi(r,t))\nonumber\\
    &=&\frac{\d R(t)}{\d t}[\bar\phi_{\l}^\L(R(t))-\bar\phi_{\g}^\L(R(t))]-[\bar J_{\l}(R(t))-\bar J_{\g}(R(t))]+\bar\upsilon\int_0^\infty \d rr^2s(\bar\phi(r,t)).
    \label{eq:R_deriv_2}
\end{eqnarray}
where in the last equality we recalled the fast-surface-thermodynamics boundary condition, $\bar\phi_\alpha(R,t)=\bar\phi_\alpha^\L(R)$ ($\alpha=\l,\g$), and $\bar J_\alpha(R)$ are the outward steady fluxes on either side of the interface of a droplet whose radius is $R$. Upon comparing Eqs.~\eqref{eq:R_deriv_1} and~\eqref{eq:R_deriv_2}, we cancel out the reaction term $\bar\upsilon\int_0^\infty \d rr^2s(\bar\phi(r,t))$ and isolate $\d R(t)/\d t$
\begin{equation}
    [\bar\phi_{\l}^\L(R(t))-\bar\phi_{\g}^\L(R(t))]\frac{\d R(t)}{\d t}=\bar J_{\l}(R(t))-\bar J_{\g}(R(t)).\label{eq:radius_growth_start}
\end{equation}
This is Eq.~(3) of the main text. To be able to compute $\d R/\d t$, we must evaluate the steady interface fluxes from the field-theory equations.

\subsection{Interim simplifications}\label{sec:simp}

In this work, we consider a regime where the deviations of the volume fraction fields from their thermodynamic-equilibrium values are assumed to be small, so linearizing Eq.~\eqref{eq:phi_dynamics} deep in each bulk might be justified. This strategy is often employed in the literature~\cite{BrayAP94,ZwickerPRE15}, as it allows one to find simple explicit expressions for the fluxes. While this strategy might be justified for the dilute phase, we encounter various destabilizing phenomena in the dense phase rendering the linearization inaccurate. Thus, for completeness, we now show how linearization is performed in a general case. We then perform it explicitly for the dilute phase in Sec.~\ref{sec:gas} for which it is valid; with it, we find the dilute-phase fluxes. Afterwards, in Sec.~\ref{sec:liquid} we perform the linearization in the dense phase, identify the problematic dense-phase phenomena, and then present the approach we use in practice to compute the dense-phase flux without linearization in Sec.~\ref{sec:liq_nonline}.

The linearization is based on the assumption that the volume fractions fields $\{\bar\phi_\alpha(r;R)\}$ remain close to their coexistence values, $\{\bar\phi_\alpha^\L(R)\}$, in each phase $\alpha=\l,\g$. This allows us to write
\begin{equation}
    \bar\phi_\alpha(r;R)\equiv\bar\phi_\alpha^\L(R)+\delta\bar\phi_\alpha(r;R).\label{eq:linearize_volume}
\end{equation}
For brevity, we will omit the reminder of the parametric dependence on $R$ henceforth. Note that, in accord with Sec.~\ref{sec:nonsticky}, the volume fractions $\phi_{\s\l}^\L$ ($\simeq1$), $\phi_{\s\g}^\L$ ($\ll1$), and $\phi_{\ns\l}^\L$ ($\to0$) are known explicitly, whereas $\phi_{\ns\g}^\L$ is still unknown. As mentioned there and in the main text, $\phi_{\ns\g}^\L$ will be determined by equating the two equations for $\d R/\d t$, obtained from the sticky and nonsticky fluxes (Eq.~\eqref{eq:radius_growth_start} or Eq.~(3) of the main text; this is done in Sec.~\ref{sec:growth} below). To obtain the equation for $\d R/\d t$, we regard $\phi_{\ns\g}^\L$ as a known parameter, and express the flux $J_{\ns}$ in term of $\phi_{\ns\g}^\L$. The linearization of Eq.~\eqref{eq:linearize_volume} is thus an ansatz for $\phi_{\ns\g}^\L$\,---\,that $\phi_{\ns\g}^\L\simeq \varphi_\ns^\infty$\,---\,which we make to be able to compute the flux $J_{\ns}$ in terms of $\phi_{\ns\g}^\L$. Once all calculations are complete and $\phi_{\ns\g}^\L$ is found, we confirm that indeed the resulting $\phi_{\ns\g}^\L$ is close to $\varphi_\ns^\infty$.

Since gradients of $\delta\bar\phi_\alpha(r)$ are over large distances of order the radius and the reaction-diffusion lengthscale, we ignore the short-distance surface tension term $\kappa_\s\nabla^2\bar\phi_\alpha$ and obtain the quasistatic linearized equations for $\delta\bar\phi_\alpha$,
\begin{equation}
    \frac{\bbar D_\alpha}{r^2}\frac\partial{\partial r}\left[r^2\frac{\partial\delta\bar\phi_\alpha(r)}{\partial r}\right]=-\bar\upsilon\left[s_\alpha+\bar k_\alpha^{\T}\delta\bar\phi_\alpha(r)\right],\label{eq:linear_react-diffuse}
\end{equation}
where we have defined an effective diffusion matrix, zeroth-, and first-order reaction rates, all evaluated at the coexistence volume fractions of phase $\alpha$,
\begin{subequations}
\begin{eqnarray}
    \bbar D_\alpha&=&\bbar\Gamma(\bar\phi_\alpha^\L)\frac{\partial^2f}{\partial\bar\phi\partial\bar\phi}\biggr|_{\bar\phi_\alpha^\L},\label{eq:diffuse_matrix}\\
    s_\alpha&=&s(\bar\phi_\alpha^\L),\label{eq:zeroth}\\
    \bar k_\alpha&=&\left.\frac{\partial s}{\partial\bar\phi}\right|_{\bar\phi_\alpha^\L}.\label{eq:first}
\end{eqnarray}
\end{subequations}
To gain intuition about these quantities, we will occasionally recall $f(\bar\phi)$ and $\{\bar\phi_\alpha^\L\}$ from the FH free energy (Sec.~\ref{sec:nonsticky}), and consider the simplest mobility matrix~\cite{KramerPOLY84,KehrPRB89,MaoSM19}, 
\begin{equation}
    \bbar\Gamma(\bar\phi)=\left(\begin{array}{cc}
        \Gamma\phi_\s & 0 \\
        0 & \Gamma\phi_\ns
    \end{array}\right).\label{eq:Gamma_mobility}
\end{equation}
A common mobility constant $\Gamma$ is taken for both the sticky and nonsticky particles since they are structurally similar, and hence the viscous resistance of the host (inert) fluid to their diffusion should be comparable.

We decouple Eqs.~\eqref{eq:linear_react-diffuse} by defining the fields
\begin{equation}
    \left(\begin{array}{c}
        \delta\Phi_\alpha(r)\\
        \delta\Psi_\alpha(r) 
    \end{array}\right)\equiv\left(\begin{array}{c}
        \bar\upsilon_\perp^\T\bbar D_\alpha\\
        \bar k_\alpha^\T
    \end{array}\right)\delta\bar\phi_\alpha(r),\label{eq:PhiPsi}
\end{equation}
They obey
\begin{subequations}
\label{eq:linear_react-diffuse_decouple}
\begin{eqnarray}
    \frac1{r^2}\frac\partial{\partial r}\left[r^2\frac{\partial\delta\Phi_\alpha(r)}{\partial r}\right]&=&0,\\
    \frac1{r^2}\frac\partial{\partial r}\left[r^2\frac{\partial\delta\Psi_\alpha(r)}{\partial r}\right]&=&\frac{s_\alpha+\delta\Psi_\alpha(r)}{\ell_\alpha^2},\label{eq:linear_react-diffuse_Psi}
\end{eqnarray}
\end{subequations}
where we defined the reaction-diffusion lengthscale in each phase,
\begin{equation}
    \frac1{\ell_\alpha^2}\equiv-\bar k_\alpha^\T\bbar D_\alpha^{-1}\bar\upsilon.\label{eq:ell}
\end{equation}
In the coming sections, we solve Eqs.~\eqref{eq:linear_react-diffuse_decouple} and discuss the choice of sign in Eq.~\eqref{eq:ell}.

\subsection{Gas phase}\label{sec:gas}

Since the gas is dilute, we consider a diagonal matrix $D_{ij\g}=\delta_{ij}D_\g$, where $D_\g$ is the effective single-particle diffusion constant that appears in the main text. This is reasonable since both the hydrodynamic and thermodynamic couplings are weak as the sticky molecules rarely encounter each other. This is further supported by inserting the FH free energy Eq.~\eqref{eq:FHfree_energy_3} and the mobility Eq.~\eqref{eq:Gamma_mobility} in Eq.~\eqref{eq:diffuse_matrix},
\begin{equation}
    \bbar D_\g=\Gamma\left(\begin{array}{cc}
        1-(2\chi-1)\left(e^{-\chi}+\frac\lambda R\right) & e^{-\chi}+\frac\lambda R \\
        \phi_{\ns\g}^\L & 1+\phi_{\ns\g}^\L
    \end{array}\right),
\end{equation}
where we have kept leading-order corrections in small $e^{-\chi}\ll1$, $\lambda/R\ll1$, and $\phi_{\ns\g}^\L\ll1$ for demonstration. Indeed, the leading order is a diagonal diffusivity matrix, $D_{ij\g}=\delta_{ij}D_\g$, with $D_\g=\Gamma$.

We proceed to investigate the reactions, $\partial \bar\phi(t)/\partial t=\bar\upsilon s(\bar\phi(t))$. Once a total amount of material is imposed (\textit{e.g.}, in our case, $\bar\upsilon_\perp^\T\bar\varphi^\infty=\varphi^\infty$), chemical equilibrium is determined by the condition $s(\bar\varphi^\infty)=0$. For it to be a stable equilibrium, we require that the Jacobian, $\bar\upsilon(\partial s/\partial\bar\phi)^\T|_{\bar\varphi^\infty}\simeq\bar\upsilon\bar k_\g^\T$ does not have any positive eigenvalues; otherwise, if the volume fractions deviate from $\bar\varphi^\infty$ (which is indeed the case around a surface, where the volume fractions are slightly different, $\bar\phi_\g^\L$), the entire dilute phase would destabilize due to the mere presence of reactions. We make the reasonable assumption that the reactions in isolation are indeed stable. The Jacobian's eigenvalues are $0$ (corresponding to steady state) and $\bar k_\g^\T\bar\upsilon$, so we require $\bar k_\g^\T\bar\upsilon<0$ (corresponding to a decaying mode). Since $\bbar D_\g$ is to leading order diagonal, we confirm that $\ell^2_\g\simeq-D_\g\bar k_\g^\T\bar\upsilon>0$ (as $D_\g>0$), so the choice of sign in Eq.~\eqref{eq:ell} is justified. 

\subsubsection{Solution of the linearized equations}\label{sec:gas_linear}

With the above preparations, we proceed to solve Eqs.~\eqref{eq:linear_react-diffuse_decouple}. We recall the boundary conditions for the gas phase\,---\,there is no deviation from coexistence at the interface, and far from the droplet the volume fractions account for the overall saturation $\bar\upsilon_\perp^\T\bar\varphi^\infty=\varphi^\infty$ and also equilibrate the reactions ($s_\g(\bar\varphi^\infty)=s_\g+\bar k_\g^\T\delta\bar\phi_\g(\infty)=0$),
\begin{subequations}
\begin{eqnarray}
    \delta\Phi_\g(R)&=&0,\qquad
    \delta\Phi_\g(\infty)=\bar\upsilon_\perp^\T\bbar D_\g\delta\bar\phi_\g(\infty),\\
    \delta\Psi_\g(R)&=&0,\qquad
    \delta\Psi_\g(\infty)=\bar k_\g^\T\delta\bar\phi_\g(\infty)=-s_\g.
\end{eqnarray}
\end{subequations}
The solutions to Eqs.~\eqref{eq:linear_react-diffuse_decouple} read
\begin{subequations}
\label{eq:PhiPsi_gas}
\begin{eqnarray}
    \delta\Phi_\g(r)&=&\bar\upsilon_\perp^\T\bbar D_\g\delta\bar\phi_\g(\infty)\left(1-\frac Rr\right),\\
    \delta\Psi_\g(r)&=&\bar k_\g^\T\delta\bar\phi_\g(\infty)\left[1-\frac{R}{r}e^{-(r-R)/\ell_\g}\right].
\end{eqnarray}
\end{subequations}
Indeed, if the deviations $\delta\phi_{i\g}(\infty)=\varphi_i^\infty-\phi_{i\g}^\L$ are small, the profiles $\delta\Phi_\g(r)$ and $\delta\Psi_\g(r)$ are also small everywhere in the dilute phase, justifying the linearization.

We prepare the mathematical property
\begin{equation}
    \bbar D_\alpha\left(\begin{array}{c}
        \bar\upsilon_\perp^\T\bbar D_\alpha\\
        \bar k_\alpha^\T
    \end{array}\right)^{-1}\left(\begin{array}{c}
        0\\
        1
    \end{array}\right)=-\ell_\alpha^2\bar\upsilon,\label{eq:matrix_property}
\end{equation}
where we have kept in mind $\bar\upsilon=(-|\upsilon_\s|,1)^\T$, $\bar\upsilon_\perp=(1,|\upsilon_\s|)^\T$, and the definition of $\ell_\alpha^2$ (Eq.~\eqref{eq:ell}). Combining it with the definitions in Eq.~\eqref{eq:PhiPsi}, we find the fluxes at the interface
\begin{equation}
    \bar J_\g(R)=-\bbar D_\g\left.\frac{\d\bar\phi_g}{\d r}\right|_R=-\bbar D_\g\left(\begin{array}{c}
        \bar\upsilon_\perp^\T\bbar D_\g\\
        \bar k_\g^\T
    \end{array}\right)^{-1}\left.\frac\d{\d r}\left(\begin{array}{c}
        \delta\Phi_\g\\
        \delta\Psi_\g
    \end{array}\right)\right|_R=-\frac{D_\g}R(\bar\varphi^\infty-\bar\phi_\g^\L)-\bar\upsilon s_\g\ell_\g.\label{eq:flux_gas}
\end{equation}
The first term corresponds to the passive diffusive influx of material from a supersaturated environment, where the total amount of material joining the sphere is proportional to the radius, $R^2J_{i\g}(R)\sim D_\g R$ ($i=\s,\ns$). The second term is the additional flux due to reactions in the dilute phase; since reactions equilibrate past distance $\ell_\g$, the amount of material joining due to reactions with rate $s_\g$ is proportional to the shell volume surrounding the interface, $R^2\bar J_\g(R)\sim s_\g R^2\ell_\g$. 

\subsubsection{Simpler main-text linearization}\label{sec:gas_example}

The results of the previous section stem from a general linearization procedure, and they will be used in Sec.~\ref{sec:growth} below. Owing to the gas phase being dilute, the results of the main text were based on a slightly simpler procedure as follows: The volume fraction fields $\bar\phi_\g(r)$ vary between (the small) $\bar\phi_\g^\L$ and (the small) $\bar\varphi^\infty$, so they are small for all $r$. Thus, we expand the reaction rates around zero volume fraction (as opposed to around $\bar\phi_\g^\L$ as above), $s(\bar\phi)\simeq s(\bar0)+(\partial s/\partial\bar\phi)^\T|_{\bar0}\bar\phi$. There are no reactions in the absence of material, so $s(\bar0)=0$. Next, we assume that there is a first-order reaction step in the interconversion mechanism, such that neither component of $(\partial s/\partial\bar\phi)^\T|_{\bar0}$ is zero. Hence, we neglect higher-order terms in the above expansion. For first-order reactions to take place, the stoichiometry must be one to one, $\bar\upsilon=(-1,1)^\T$. Furthermore, we denote as in the main text, $(\partial s/\partial\bar\phi)|_{\bar0}=(k_{\s\to\ns},-k_{\ns\to\s})^\T$ (with constant $k_{\s\to\ns},k_{\ns\to\s}>0$), where the sign is such that $\s$ is lost when creating $\ns$ and vice versa. Indeed, $k_{i\to j}$ are uniquely determined from the full reaction mechanism $s(\bar\phi)$. Overall,
\begin{equation}
    s(\bar\phi_\g(r))\simeq k_{\s\to\ns}\phi_{\s\g}(r)-k_{\ns\to\s}\phi_{\ns\g}(r),\label{eq:gas_mechanism}
\end{equation}
which was used for the gas-phase reactions in the main text. This approximation simply states that only the first-order reactions occur with a meaningful rate, as encounters between molecules required for higher-order reactions are too rare. This need not imply necessarily that Eq.~\eqref{eq:gas_mechanism} is the full mechanism; a different limit of the full mechanism might apply in the liquid phase. 

In the language of the more general linearization of Sec.~\ref{sec:simp}, the main-text linearization amounts to choosing $\bar k_\g=(k_{\s\to\ns},-k_{\ns\to\s})^\T$ and $s_\g=k_{\s\to\ns}\phi_{\s\g}^\L-k_{\ns\to\s}\phi_{\ns\g}^L$. Using the diagonal diffusivity $D_{ij\g}=\delta_{ij}D_\g$ and the stoichiometry $\bar\upsilon=(-1,1)^\T$, we obtain $1/\ell_\g^2=(|\bar\upsilon_\s|k_{\s\to\ns}+k_{\ns\to\s})/D_\g$. Upon inserting these facts in Eqs.~\eqref{eq:PhiPsi_gas}, we find
\begin{subequations}
\begin{eqnarray}
    \delta\phi_{\s\g}(r)+\delta\phi_{\ns\g}(r)&=&(\varphi^{\infty}-\phi_{\s\g}^{\L}-\phi_{\ns\g}^{\L})\left(1-\frac Rr\right),\\
    k_{\s\to\ns}\delta\phi_{\s\g}(r)-k_{\ns\to\s}\delta\phi_{\ns\g}(r)&=&-(k_{\s\to\ns}\phi_{\s\g}^{\L}-k_{\ns\to\s}\phi_{\ns\g}^{\L})\left[1-\frac{R}{r}e^{-(r-R)/\ell_\g}\right],\quad
\end{eqnarray}
\end{subequations}
Explicitly,
\begin{subequations}
\begin{eqnarray}
    \phi_{\s\g}(r)&=&\varphi_\s^\infty-\left[\varphi_\s^\infty-\frac{k_{\ns\to\s}(\phi_{\s\g}^{\L}+\phi_{\ns\g}^{\L})}{k_{\s\to\ns}+k_{\ns\to\s}}\right]\frac{R}{r}+\frac{k_{\s\to\ns}\phi_{\s\g}^{\L}-k_{\ns\to\s}\phi_{\ns\g}^{\L}}{k_{\s\to\ns}+k_{\ns\to\s}}\frac{R}{r}e^{-(r-R)/\ell_\g},\quad\\
    \phi_{\ns\g}(r)&=&\varphi_\ns^\infty-\left[\varphi_\ns^\infty-\frac{k_{\s\to\ns}(\phi_{\s\g}^{\L}+\phi_{\ns\g}^{\L})}{k_{\s\to\ns}+k_{\ns\to\s}}\right]\frac{R}{r}-\frac{k_{\s\to\ns}\phi_{\s\g}^{\L}-k_{\ns\to\s}\phi_{\ns\g}^{\L}}{k_{\s\to\ns}+k_{\ns\to\s}}\frac{R}{r}e^{-(r-R)/\ell_\g}.\quad
\end{eqnarray}
\end{subequations}
where $\bar\varphi^\infty=(k_{\ns\to\s}\varphi^\infty/(k_{\s\to\ns}+k_{\ns\to\s}),k_{\s\to\ns}\varphi^\infty/(k_{\s\to\ns}+k_{\ns\to\s}))^\T$. These are the profiles sketched in Fig.~2 of the main text. As mentioned earlier, the first two terms are the standard diffusive profile around a droplet, whereas the last terms oppose departure from the chemical equilibrium, increasing the diffusive flux out of small droplets and into large droplets. To see that explicitly, the fluxes from Eq.~\eqref{eq:flux_gas} simplify as
\begin{subequations}
\begin{eqnarray}
    J_{\s\g}(R)&=&-\frac{D_\g}R(\varphi_\s^\infty-\phi_{\s\g}^{\L})+\frac{D_\g}\ell\frac{k_{\s\to\ns}\phi_{\s\g}^{\L}-k_{\ns\to\s}\phi_{\ns\g}^{\L}}{k_{\s\to\ns}+k_{\ns\to\s}},\\
    J_{\ns\g}(R)&=&-\frac{D_\g}R(\varphi_\ns^\infty-\phi_{\ns\g}^{\L})-\frac{D_\g}\ell\frac{k_{\s\to\ns}\phi_{\s\g}^{\L}-k_{\ns\to\s}\phi_{\ns\g}^{\L}}{k_{\s\to\ns}+k_{\ns\to\s}}.
\end{eqnarray}
\end{subequations}
which coincide with Eq.~(6) of the main text.

In the derivations to follow, we use the more general linearized result of Sec.~\ref{sec:gas_linear}. It accounts for cases where a higher-order reaction dominates, namely, if $(\partial s/\partial\bar\phi)|_{\bar0}=\bar0$. For example, $s(\bar\phi)=k_{\s\to\ns}\phi_\s^2-k_{\ns\to\s}\phi_\ns^2$ with $\bar\upsilon=(-1,1)$, which describes a reaction mechanism where a collision between two particles of the same form is needed to form two of the other form. In that case, one should have continued the above series to $s(\bar\phi)\simeq \bar\phi^\T(\partial^2 s/\partial\bar\phi\partial\bar\phi^\T)|_{\bar0}\bar\phi$, such that $\bar k_\g^\T=2\bar\phi_\g^{\L\T}(\partial^2 s/\partial\bar\phi\partial\bar\phi^\T)|_{\bar0}$ and $s_\g=\bar\phi_\g^{\L\T}(\partial^2 s/\partial\bar\phi\partial\bar\phi^\T)|_{\bar0}\bar\phi_\g^L=(1/2)\bar k_\g^\T\bar\phi_\g^\L$ (as opposed to the above, effectively first-order treatment where $s_\g=\bar k_\g^\T\bar\phi_\g^\L$). In total, the general linearization of Sec.~\ref{sec:simp} accounts for non-first-order reaction mechanisms, which is a more general gas-phase formalism than the one in the main text.  

\subsection{Liquid phase}\label{sec:liquid}

In the liquid phase, the diffusion matrix is not diagonal. For example, for the FH free energy,
\begin{equation}
    \bbar D_\l=\Gamma \left(\begin{array}{cc}
        e^\chi(1-2\chi e^{-\chi}+\phi_{\ns\g}^\circ) & e^\chi(1-e^{-\chi}+\phi_{\ns\g}^\circ) \\
        0 & 1
    \end{array}\right),\label{eq:diff_liq_FH}
\end{equation}
where we have kept up to first-order corrections in small $e^{-\chi}\ll1$ and $\phi_{\ns\g}^\circ\ll1$ and we omitted Laplace-pressure corrections for simplicity. This matrix incorporates both the mobility and the free-energy response to fluctuations in the volume fraction fields. Since the liquid phase is dense in sticky protein, the sticky proteins quickly ($\sim\Gamma e^\chi$) rearrange in response to a perturbation in either the sticky or nonsticky volume fraction. (At the same time, the nonsticky protein is inert, so $D_{\ns\s\l}\sim0$ and $D_{\ns\ns\l}\sim1$ as if it is still an ideal gas.) This dominant order-$e^\chi$ term arises from the second-order derivative of the entropy contribution to the FH free energy, as the sticky volume fraction is very close to $1$ (within $\sim e^{-\chi}$) and so approaching it further is entropically costly. Interestingly, since the sticky proteins attract each other, a sticky fluctuation is slightly more stable than a nonsticky fluctuation (\textit{cf.} $D_{\s\s\l}-D_{\s\ns\l}=-\Gamma(2\chi-1)<0$ with $\chi\gg1$) as the former is slightly stabilized by the energy term. This will play a role in determining $\ell_\l^2$ in Sec.~\ref{sec:liq_example2}. Generally, the liquid-phase analogue for the reaction-stability criterion ($\bar k_\l^\T\bar\upsilon<0$) does not impose anything on $\ell_\l^2=-\bar k_\l^\T\bbar D_\l^{-1}\bar\upsilon$ since $\bbar D_\l$ is not diagonal. We proceed to Secs.~\ref{sec:liq_lin} and~\ref{sec:liq_example1} as if the sign convention in Eq.~\eqref{eq:ell} is correct; this will not be the case in Sec.~\ref{sec:liq_example2}.

\subsubsection{Solution of the linearized equations}\label{sec:liq_lin}

We recall the boundary conditions for the liquid phase\,---\,no deviation from coexistence at the interface, and simply nondiverging values at the droplet center,
\begin{subequations}
\label{eq:liquid_BC}
\begin{eqnarray}
    \delta\Phi_\l(R)&=&0,\qquad
    \delta\Phi_\l(0)<\infty,\\
    \delta\Psi_\l(R)&=&0,\qquad
    \delta\Psi_\l(0)<\infty.\label{eq:liquid_BC_Psi}
\end{eqnarray}
\end{subequations}
The solutions to Eqs.~\eqref{eq:linear_react-diffuse_decouple} read
\begin{subequations}
\label{eq:PhiPsi_liq}
\begin{eqnarray}
    \delta\Phi_\l(r)&=&0,\\
    \delta\Psi_\l(r)&=&-s_\l\left[1-\frac{R}{r}\frac{\sinh(r/\ell_\l)}{\sinh(R/\ell_\l)}\right].\label{eq:Psi_liq}
\end{eqnarray}
\end{subequations}
In Secs.~\ref{sec:liq_example1} and~\ref{sec:liq_example2}, we examine the validity of the linearization approximation, and find two destabilizing effects that one must avoid in order to consider the possibility of acceleration.

Supposing momentarily that the linearization is valid, using the relation in Eq.~\eqref{eq:matrix_property} and the definitions in Eq.~\eqref{eq:PhiPsi}, we find the fluxes at the interface
\begin{equation}
    \bar J_\l(R)=-\bbar D_\l\left.\frac{\d\bar\phi_\l}{\d r}\right|_R=-\bbar D_\l\left(\begin{array}{c}
        \bar\upsilon_\perp^\T\bbar D_\l\\
        \bar k_\l^\T
    \end{array}\right)^{-1}\left.\frac\d{\d r}\left(\begin{array}{c}
        \delta\Phi_\l\\
        \delta\Psi_\l
    \end{array}\right)\right|_R=\bar\upsilon s_\l\ell_\l\left[\coth\left(\frac R{\ell_\l}\right)-\frac{\ell_\l}R\right].\label{eq:flux_liq}
\end{equation}
Since the droplet is finite sized, in contrast to the gas phase, the reactions might not equilibrate at the core. This manifests as follows: If the droplet is smaller than the inter-reaction distance of the liquid phase, $R\ll\ell_\l$, then the volume fractions are almost uniform everywhere inside the droplet, and the material is uniformly converted with rate $s_\l$ throughout the entire droplet. Thus, the flux is proportional to the volume of the droplet, $R^2\bar J_\l(R)\sim s_\l R^3$. Explicitly,
\begin{equation}
    \bar J_\l(R\ll\ell_\l)\simeq \frac13\bar\upsilon s_\l R.\label{eq:sl_small}
\end{equation}
In the other limit, when the droplet is sufficiently large $R\gg\ell_\l$, the particles will typically have undergone multiple interconversion reactions past a distance $\ell_\l$ into the droplet, and will have reached chemical equilibrium. Thus, net interconversions occur only in a shell of this width, $R^2\bar J_\l(R)\sim s_\l R^2\ell_\l$. Explicitly,
\begin{equation}
    \bar J_\l(R\gg\ell_\l)\simeq \bar\upsilon s_\l \ell_\l\left(1-\frac{\ell_\l}R\right).\label{eq:sl_large}
\end{equation}

\subsubsection{Example: Crossing the spinodal}\label{sec:liq_example1}

We consider the simplest conversion reaction mechanism, $s(\bar\phi)=k_{\s\to\ns}\phi_\s-k_{\ns\to\s}\phi_\ns$ (with $\bar\upsilon=(-1,1)^\T$ and $\bar\upsilon_\perp=(1,1)^\T$) with the FH free energy. In this case, $s_\l=k_{\s\to\ns}+\mathcal{O}(e^{-\chi})$, $\bar k_\l=(k_{\s\to\ns},-k_{\ns\to\s})^\T$, and upon inverting the nondiagonal diffusivity of Eq.~\eqref{eq:diff_liq_FH}, $1/\ell_\l^2=(k_{\s\to\ns}+k_{\ns\to\s})/\Gamma+\mathcal{O}(\chi e^{-\chi})$.
Inserting these expressions in Eq.~\eqref{eq:PhiPsi_liq}, and inverting Eq.~\eqref{eq:PhiPsi} using the property of Eq.~\eqref{eq:matrix_property}, we find
\begin{eqnarray}
\label{eq:phi_liq1} 
    \bar\phi_\l(r)&=&\bar\phi_\l^\L+k_{\s\to\ns}\ell_\l^2\bbar{D}_\l^{-1}\bar\upsilon\left[1-\frac{R}{r}\frac{\sinh(r/\ell_\l)}{\sinh(R/\ell_\l)}\right]\nonumber\\&=&\bar\phi_\l^\L+\frac{k_{\s\to\ns}}{k_{\s\to\ns}+k_{\ns\to\s}}\left(\begin{array}{c}
        -1\\
        1 
    \end{array}\right)\left[1-\frac{R}{r}\frac{\sinh(r/\ell_\l)}{\sinh(R/\ell_\l)}\right].
\end{eqnarray}

In contrast to the gas phase (Sec.~\ref{sec:gas_example}), $s_\l\sim 1\gg s_\g\sim\bar\phi_{\g}^\L$. As a result, we see in Eq.~\eqref{eq:phi_liq1} that the deviations from the coexistence (binodal) volume fractions $\bar\phi_\l^\L$ might be of order up to $k_{\s\to\ns}/(k_{\s\to\ns}+k_{\ns\to\s})\sim1$. When the droplets are small ($R\ll\ell_\l$), the second term would be of order $[k_{\s\to\ns}/(k_{\s\to\ns}+k_{\ns\to\s})](R^2-r^2)/\ell_\l^2$ which is small ($R^2/\ell_\l^2\ll1$), so the linearization approach is indeed justified. On the other hand, for a large droplet, the volume fractions at the core are modified by the above factor $k_{\s\to\ns}/(k_{\s\to\ns}+k_{\ns\to\s})$. Therefore, if reactions favor the formation of nonsticky species ($k_{\s\to\ns}>k_{\ns\to\s}$, which is what is needed to obtain a meaningful acceleration), the second term in Eq.~\eqref{eq:phi_liq1} originating from the reactions becomes of order $\sim1$. This (i) invalidates the linearization approach, and (ii) even worse, suggests that if reactions dominate over the compressibility, this might drive the volume fractions beyond the spinodal, causing the whole dense-phase to destabilize. We recall that in this situation, where $s_\l\sim \Gamma/\ell_\l^2\gg \Gamma(\varphi_\s^\infty-\phi_{\s\g}^\L)/\ell_\l^2$ (where $\Gamma=D_\g$), we concluded in the main text that size control takes place~\cite{ZwickerPRE15}, and hence this instability is averted. Although, no acceleration would be achieved in this case.

\subsubsection{Example: Turing instability}\label{sec:liq_example2}

To potentially avoid size control, we consider another toy mechanism which suppresses reactions inside the droplet: $s(\bar\phi)=(1-\phi_\s-\phi_\ns)(k_{\s\to\ns}\phi_\s-k_{\ns\to\s}\phi_\ns)$ with $\bar\upsilon=(-1,1)^\T$. It might describe a situation where a kinase and phosphatase do not have as much access to the dense phase as they have to the dilute phase. Instead, being a part of the cell's proteome, they only have the available volume fraction $\phi_\o=1-\phi_\s-\phi_\ns$ inside droplets. (Of the proteome, they constitute some fraction which is absorbed into the rates $k_{\s\to\ns}$ and $k_{\ns\to\s}$.) Combining, to convert $i\to j$ with rate $k_{i\to j}$, we need an encounter between the right enzyme (available volume fraction $1-\phi_\s-\phi_\ns$) and the substrate ($\phi_i$). (This mechanism is indeed chemically stable for all volume fractions, $(\partial s/\partial\bar\phi)^\T\bar\upsilon=-(k_{\s\to\ns}+k_{\ns\to\s})(1-\phi_\s-\phi_\ns)<0$, where $\phi_\o=1-\phi_\s-\phi_\ns>0$).) In the gas phase, this mechanism reverts to Eq.~\eqref{eq:gas_mechanism} (since $\phi_{\s\g}+\phi_{\ns\g}\ll1$).
By construction, this mechanism suppresses the total zeroth-order reaction rate in the dense phase: To leading order in small $e^{-\chi}$, it reads, $s_\l=k_{\s\to\ns}e^{-\chi}$, \textit{i.e.}, sticky proteins are converted to nonsticky ones as if the whole droplet consists of only sticky proteins (nearly-$1$ volume fraction), though the enzymes only have a small available volume fraction ($e^{-\chi}$) to access the droplet. Furthermore, the linear corrections to the zeroth-order reaction rate due to fluctuations in the sticky and nonsticky volume fractions read $\bar k_\l=(-k_{\s\to\ns}+2k_{\s\to\ns}e^{-\chi},-k_{\s\to\ns}+(k_{\s\to\ns}-k_{\ns\to\s})e^{-\chi})^\T$. That is, $s_\l$ is an overestimate of how much sticky material is being converted to nonsticky; due to the strong exclusion, once there is a downward fluctuation in either $\s$ or $\ns$ volume fraction, the enzymes would have extra volume to enter the droplet and create back sticky particles with a rate $k_{\s\to\ns}$. 

These expressions, combined with the diffusion matrix of Eq.~\eqref{eq:matrix_property}, are inserted in Eq.~\eqref{eq:ell} yield $1/\ell_\l^2=e^{-\chi}[(k_{\s\to\ns}+k_{\ns\to\s})/\Gamma][1-2\chi k_{\s\to\ns}/(k_{\s\to\ns}+k_{\ns\to\s})]$. This expression can be understood as follows: First, the combination of very little access for interconversion enzymes into droplets ($s\sim1-\phi_\s-\phi_\ns\ll1$) with the rapid relaxation of fluctuations in sticky volume fraction field ($D_{\s i\l}\sim e^\chi$), implies that the typical distance between reactions is large, $\ell_\l\sim e^{\chi/2}$. Second, $\ell_\l^2$ in its current definition (Eq.~\eqref{eq:ell}) might become negative if $\chi>(1+k_{\ns\to\s}/k_{\s\to\ns})/2$, \textit{i.e.}, if the nonsticky protein is not converted to the sticky form fast enough. This is a result of a Turing instability, whereby diffusion destabilizes the otherwise stable reactions ($(\partial s/\partial\bar\phi)^\T\bar\upsilon<0$). Specifically, as we identified in Eq.~\eqref{eq:diff_liq_FH}, the entropic response of the sticky volume-fraction field to a fluctuation in the nonsticky one is faster than the response to a fluctuation in the sticky one (since the latter is slightly stabilized energetically, parametrized with $\chi$). Thus, unless nonsticky proteins are converted back fast enough, the sticky proteins would separate from regions where nonsticky proteins are created. Since surface thermodynamics maintain $\phi_{\s\l}(R)=\phi_{\s\l}^\L$ at the interface, this instability would start at the core of the droplet every distance ``$\mathrm i\ell_\l$''. Due to the surface tension, we might expect that this would culminate in bubbles~\cite{TjhungPRX18}. 

Mathematically, Turing instability manifests like so: First, we amend Eq.~\eqref{eq:ell} to $\ell_\T=\mathrm{i}\ell_\l$,
\begin{equation}
    \frac1{\ell_\T^2}\equiv\bar k_\l^\T\bbar D_\l^{-1}\bar\upsilon.
\end{equation}
The linearized Eq.~\eqref{eq:linear_react-diffuse_Psi} modifies to
\begin{equation}
    \frac1{r^2}\frac\partial{\partial r}\left[r^2\frac{\partial\delta\Psi_\l(r)}{\partial r}\right]=-\frac{s_\l+\delta\Psi_\l(r)}{\ell_\T^2}.
\end{equation}
With the boundary conditions of Eq.~\eqref{eq:liquid_BC_Psi}, the previous solution (Eq.~\eqref{eq:Psi_liq}) becomes instead
\begin{equation}
    \delta\Psi_\l(r)=-s_\l\left[1-\frac{R}{r}\frac{\sin(r/\ell_\T)}{\sin(R/\ell_\T)}\right].\label{eq:Psi_liq_Turing}
\end{equation}
Indeed, when the droplets are larger than the Turing length, $R>\pi\ell_\T$, the instabilities start forming, and according to Eq.~\eqref{eq:Psi_liq_Turing} the volume-fraction fields found from the linear theory exhibit divergence. As in Sec.~\ref{sec:liq_example1}, this indicates that the linear theory is no longer valid, and the volume fraction fields might cross the spinodal to form ``dilute pockets'' inside the droplets, whose wavelength is $1/\ell_\T$. 

However, for small droplets ($R\ll\ell_\T$), the patterns are yet to develop, and instead the volume fractions are roughly uniform inside the droplet, as in Sec.~\ref{sec:liq_lin}. Replacing $-\ell_\l^2\to\ell_\T^2$ in Eq.~\eqref{eq:matrix_property}, the previously-obtained flux (Eq.~\eqref{eq:flux_liq}) becomes instead
\begin{equation}
    \bar J_\l(R)=\bar\upsilon s_\l\ell_\T\left[\frac{\ell_\T}R-\cot\left(\frac R{\ell_\T}\right)\right].
\end{equation}
In the limit $R\ll\ell_\T$, we once again obtain Eq.~\eqref{eq:sl_small}, whereby the material flux is due to predominant conversion of sticky protein into nonsticky form with rate proportional to the droplet volume, $R^2\bar J_\l(R)\sim k_{\s\to\ns} R^3$.

\subsubsection{Linearization-free treatment}\label{sec:liq_nonline}

If the reactions destabilize the liquid phase (whether due to the passing of the spinodal line or through a Turing-instability mechanism), discussing accelerated ripening becomes meaningless. Thus, we assume that neither of these effects occurs, and the free energy opposes any reaction-related instabilities. With this assumption, we proceed to implicitly find the dense-phase fluxes without linearizing the volume-fraction field equations.

We revert back to the quasi-static approximation (to before Sec.~\ref{sec:simp}). With spherical symmetry, our starting point is the steady-state equation for the volume-fraction field,
\begin{equation}
    0=-\frac1{r^2}\frac\partial{\partial r}[r^2\bar J_\l(r)]+\bar\upsilon s(\bar\phi_\l(r)),
\end{equation}
where $J_\l(r)$ is the (unlinearized) steady-state flux within the dense phase for a fixed radius $R$, and the boundary conditions are given in Eqs.~\eqref{eq:liquid_BC}. We multiply both sides by $r^2$ and integrate from $0$ to $R$, which results in the general expression
\begin{equation}
    R^2\bar J_\l(R)=\bar\upsilon\int_0^R\d rr^2s(\bar\phi_\l(r)).
\end{equation}

We posit that the reactions have a certain reaction-diffusion lengthscale $\ell_\l$ setting the reaction equilibration distance inside the droplet. Since we have abandoned the linearized approach, it is not given by Eq.~\eqref{eq:ell}, and it is instead ``phenomenological''. We determine it as follows: For $R\ll\ell_\l$, the reactions do not equilibrate at the droplet core, and so we approximate $\bar\phi_\l(r)\simeq\bar\phi_\l^\L$ per the boundary conditions set at the interface. Thus,
\begin{equation}
    \bar J_\l(R\ll\ell_\l)\simeq\bar\upsilon\frac1{R^2}\int_0^R\d rr^2s(\bar\phi_\l^\L)=\frac13\bar\upsilon s_\l R,
\end{equation}
where we have naturally recovered $s_\l=s(\bar\phi_\l^\L)$. On the other hand, for $R\gg\ell_\l$, we expect the scaling $R^2\bar J_\l(R)\sim s_\l R^2\ell_\l$, as the reactions equilibrate within a distance $\ell_\l$ from the interface. Thus, we define $\ell_\l$ such that the following approximation is adequate:
\begin{equation}
    s(\bar\phi_\l(r))\simeq \begin{cases}
        s(\bar\phi_\l^\L),& R-\ell_\l<r<R,\\
        0,& 0<r<R-\ell_\l.
    \end{cases}
\end{equation}
(For instance, in the experiment of Ref.~\cite{TenaSolsonaCSC21}, $\ell_\l$ would be the distance that water molecules can penetrate into a hydrophobic condensate.) This results in
\begin{equation}
    \bar J_\l(R\ll\ell_\l)\simeq\bar\upsilon\frac1{R^2}\int_{R-\ell_\l}^R\d rr^2s(\bar\phi_\l^\L)=\frac13\bar\upsilon s_\l\frac{R^3-(R-\ell_\l)^3}{R^2}\simeq\bar\upsilon s_\l\ell_\l\left(1-\frac{\ell_\l}R\right),
\end{equation}
where the second approximation relied on $R\gg\ell_\l$. With this, we obtained Eqs.~\eqref{eq:sl_small} and~\eqref{eq:sl_large} for any reaction protocol (so long as it does not destabilize the liquid phase) without linearization. 

For compactness, we define
\begin{equation}
    \bar J_\l(R)=\bar\upsilon s_\l\ell_\g G\left(\frac R{\ell_\g},\frac{\ell_\l}{\ell_\g}\right),\label{eq:flux_liq_gen}
\end{equation}
where we defined a function that interpolates between the large- and small-droplet limits,
\begin{equation}
    G\left(\frac R{\ell_\g},\frac{\ell_\l}{\ell_\g}\right)=\begin{cases}
\frac{1}{3}\frac{R}{\ell_\g}, & R\ll\ell_\l,\\
\frac{\ell_\l}{\ell_\g}\left(1-\frac{\ell_\l}{\ell_\g}\frac{\ell_\g}R\right), & R\gg\ell_{\l}.
\end{cases}\label{eq:Ginterpolant}
\end{equation}
This is Eq.~(7) of the main text, which therefore avoids the unjustified dense-phase linearization. Instead, we have compromized for ``phenomenological'' dense-phase rate constant $s_\l$ and reactive equilibration distance $\ell_\l$.

\subsection{Droplet growth law}\label{sec:growth}

Equipped with the gas phase flux, Eq.~\eqref{eq:flux_gas} (where the linearization is adequate since the mixture is only slightly supersaturated, $0<\phi_{\s\g}^\circ\lesssim\varphi_\s^\infty\ll1$), and the liquid-phase flux, Eq.~\eqref{eq:flux_liq_gen} (which we have obtained without linearization, though we needed to assume that the reactions do not destablize the dense phase), we proceed to compute the radius growth rate, expressed from either the sticky and nonsticky fluxes. From these equations, we will extract $\phi_{\ns\g}^\L$ and the sought-after $\d R/\d t$.

Inserting Eqs.~\eqref{eq:flux_gas} and~\eqref{eq:flux_liq_gen} in Eq.~\eqref{eq:radius_growth_start}, we find
\begin{equation}
    [\bar\phi_{\l}^\L(R(t))-\bar\phi_{\g}^\L(R(t))]\frac{\d R(t)}{\d t}=\frac{D_\g}{R(t)}[\bar\varphi^\infty-\bar\phi_\g^\L(R(t))]+\bar\upsilon\ell_\g\left[ s_\g+s_\l G\left(\frac {R(t)}{\ell_\g},\frac{\ell_\l}{\ell_\g}\right)\right].\label{eq:radius_growth_explicit}
\end{equation}
The first term on the right-hand side is the contribution from passive Ostwald ripening and the second term is the consequence of reactions occurring on either side of the interface. Upon multiplying both sides of Eq.~\eqref{eq:radius_growth_explicit} by $\bar\upsilon_\perp^\T=(1,|\upsilon_\s|)$, we find Eq.~(8) of the main text in the case of general stoichiometric ratio,
\begin{equation}
    \frac{\d R(t)}{\d t}=\frac{D_\g}{R(t)}[\varphi^\infty-\phi_{\s\g}^\L(R(t))-|\upsilon_\s|\phi_{\ns\g}^\L(R(t))],\label{eq:radius_growth_conserved}
\end{equation}
where we recalled $\bar\upsilon_\perp^\T\bar\varphi^\infty=\varphi^\infty$ and kept the leading order $\bar\upsilon_\perp^\T[\bar\phi_{\l}^\L(R)-\bar\phi_{\g}^\L(R)]\simeq1$ (as $\phi_{\s\l}^\L\simeq1$, $\phi_{\s\g}^\L,\phi_{\ns\g}^\L\ll1$, and $\phi_{\ns\l}^\L\to0$). Since $\bar\upsilon_\perp^\T\bar\phi(r)$ is a conserved field which is not directly impacted by the reactions, the growth rate in term of the conserved diffusive flux adopts the functional form of passive Ostwald ripening. 

We do not know yet $\phi_{\ns\g}^\L(R)$ in Eq.~\eqref{eq:radius_growth_conserved}. By multiplying both sides of Eq.~\eqref{eq:radius_growth_explicit} by the row vector $(\phi_{\ns\g}^\L(R)-\phi_{\ns\l}^\L(R),\phi_{\s\l}^\L(R)-\phi_{\s\g}^\L(R))$ ($\simeq(0,1)$), we find to leading order in small $\phi_{\s\g}^\L,1-\phi_{\s\l}^\L,\phi_{\ns\g}^\L,\phi_{\ns\l}^\L\ll1$,
\begin{equation}
    \frac{\ell_\g}R[\varphi^\infty_\ns-\phi_{\ns\g}^\L(R)]=-\frac{\ell_\g^2}{D_\g}\left[ s_\g+s_\l G\left(\frac {R}{\ell_\g},\frac{\ell_\l}{\ell_\g}\right)\right],\label{eq:radius_growth_phi_nsg}
\end{equation}
from which $\phi_{\ns\g}^\L(R)$ will be extracted. Since $\phi_{\ns\l}^\L\to0$ is subdominant and $\phi_{\s\l}^\L\to1$ is determined independently of $\phi_{\ns\g}^\L$, neither $s_\l$ nor $\ell_\l$ depend on $\phi_{\ns\g}^\L$. Therefore, only the total zeroth-order reaction rate $s_\g=s(\phi_{\s\g}^\L,\phi_{\ns\g}^\L)$ and inter-reaction distance $\ell_\g=\{D_\g/[-\bar k_\g^\T(\phi_{\s\g}^\L,\phi_{\ns\g}^\L)\bar\upsilon]\}^{1/2}$ might depend on $\phi_{\ns\g}^\L$. (In the simpler case of the main text and Sec.~\ref{sec:gas_example}, $\ell_\g$ does not after all.) Upon determining $\phi_{\ns\g}^\L$ from Eq.~\eqref{eq:radius_growth_phi_nsg}, the radius growth law can be found explicitly from Eq.~\eqref{eq:radius_growth_conserved}. We proceed to do so in the different limits of $R$ versus $\ell_\g$ and $\ell_\l$. The physical outcomes obtained for various values of $s_\l$ and droplet size $R$ are sketched in Fig.~\ref{fig:phasediagram}.

\begin{figure}
    \centering
    \includegraphics[width=0.9\linewidth]{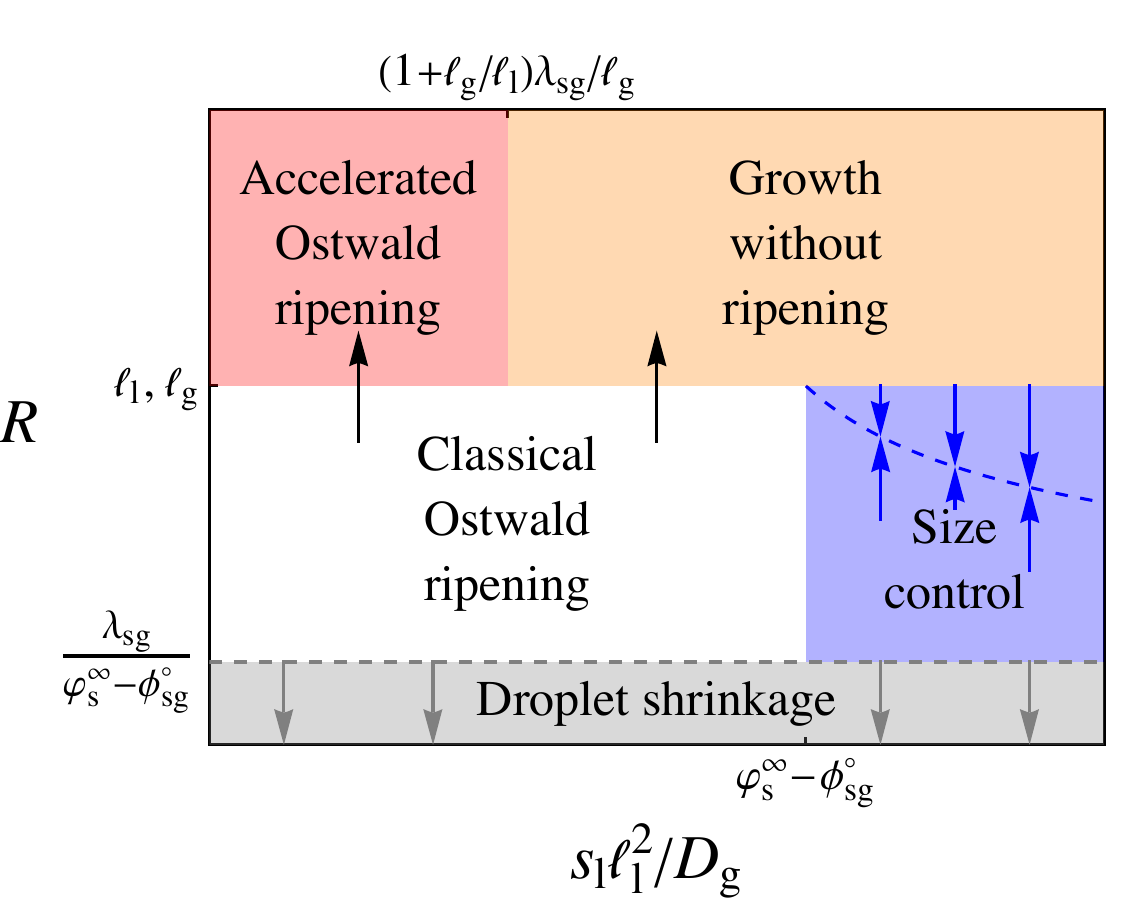}
    \caption{``Phase diagram'' of droplet dynamics as a function of total dense phase reaction rate $s_\l$ and droplet size $R(t)$. Reactions in the dilute phase may accelerate Ostwald ripening (\textit{cf}. Fig.~2 of the main text), whereas reactions in the dense phase hinder droplet growth and ripening by converting particles from sticky to nonsticky.
    Small droplets with $R\ll\lambda_{\s\g}/(\varphi_\s^\infty-\phi_{\s\g}^\circ)$ always tend to shrink due to large Laplace pressure. Other regions are approximately bounded by the inter-reaction diffusion length $\ell_\g$. For very fast dense phase reactions, $s_\l\ell_\l^2/D_\g\gg\varphi_\s^\infty-\phi_{\s\g}^\circ$, size control is inevitable with stable radius $R_\mathrm{sc}^2=3D_\g(\varphi_\s^\infty-\phi_{\s\g}^\circ)/s_\l$. Otherwise, droplets can reach the regime $R\gg\ell_\l,\ell_\g$, where: For intermediate dense phase reaction rate, $(1+\ell_\g/\ell_\l)\lambda_{\s\g}/\ell_\l\ll s_\l\ell_\l^2/D_\g\ll \varphi_\s^\infty-\phi_{\s\g}^\circ$, a negative effective surface tension prevents small droplets from shrinking. For slow dense phase reaction rate, $s_\l\ell_\l^2/D_\g\ll(1+\ell_\g/\ell_\l)\lambda_{\s\g}/\ell_\l$, Ostwald ripening occurs with an accelerated rate. Parameters: $R$, droplet radius; $s_\l$, total dense phase reaction rate; $\varphi_\s^\infty-\phi_{\s\g}^\circ$, dilute phase volume fraction relative to the coexistence value; $\lambda_{\s\g}$, Laplace-pressure correction to the volume fraction of the sticky species in the gas phase; $\ell_\l,\ell_\g$, characteristic distance between reactions in each phase; $D_\g$, dilute-phase diffusion constant.}
    \label{fig:phasediagram}
\end{figure}

\subsubsection{Small droplet}\label{sec:small_droplet}

We begin with the limit $R\ll\ell_\g,\ell_\l$. Using Eq.~\eqref{eq:Ginterpolant}, Eq.~\eqref{eq:radius_growth_phi_nsg} becomes
\begin{equation}
    \frac{\ell_\g(\phi_{\ns\g}^\L)}R(\varphi^\infty_\ns-\phi_{\ns\g}^\L)=-\frac{\ell_\g^2(\phi_{\ns\g}^\L)}{D_\g}\left[ s_\g(\phi_{\ns\g}^\L)+\frac{s_\l}{3}\frac{R}{\ell_\g(\phi_{\ns\g}^\L)}\right],\label{eq:radius_growth_phi_nsg_small}
\end{equation}
where for brevity, we again suppress the $R$-dependence of $\bar\phi_\g^\L(R)$, but bring forth the dependence of $s_\g$ and $\ell_\g$ on $\phi_{\ns\g}^\L$. (They also depend on $\phi_{\s\g}^\L$, but it is already known from the free energy; see Sec.~\ref{sec:thermo}.) Solving Eq.~\eqref{eq:radius_growth_phi_nsg_small} iteratively up to a leading-order correction for $R/\ell_\g\ll1$ yields
\begin{equation}
    \phi_{\ns\g}^\L=\varphi^\infty_\ns+\frac{\ell_\g(\phi_{\ns\g}^\L)R}{D_\g} s_\g(\phi_{\ns\g}^\L)+\frac{s_\l R^2}{3D_\g}\simeq\varphi^\infty_\ns+\frac{\ell_\g(\varphi^\infty_\ns)R}{D_\g} s_\g(\varphi^\infty_\ns)+\frac{s_\l R^2}{3D_\g}.
\end{equation}
Indeed, $\phi_{\ns\g}^\L\simeq\varphi_{\ns}^\infty$, thus justifying the linearization ansatz (Eq.~\eqref{eq:linearize_volume}) for the nonsticky species as well.

Inserting this result in Eq.~\eqref{eq:radius_growth_conserved}, we find
\begin{equation}
    \frac{\d R(t)}{\d t}=\frac{D_\g}{R(t)}[\varphi^\infty_\s-\phi_{\s\g}^\L(R(t))]+|\upsilon_\s|\ell_\g(\varphi^\infty_\ns) [-s_\g(\varphi^\infty_\ns)]-\frac{|\upsilon_\s|s_\l R(t)}{3}.\label{eq:dRdt_sizecontrol}
\end{equation}
The first term corresponds to passive Ostwald ripening, the second term is the acceleration due to reactions in the dilute phase, and the third term stems from the growth-opposing reactions in the dense droplet. We analyze size control by rewriting Eq.~\eqref{eq:dRdt_sizecontrol} as
\begin{equation}
    \frac{\d R(t)}{\d t}=\frac{D_\g}{R(t)}\left\{[\varphi^\infty_\s-\phi_{\s\g}^\L(R(t))]+|\upsilon_\s|\frac{-s_\g(\varphi^\infty_\ns)}{-\bar k_\g^\T\bar\upsilon}\frac{R(t)}{\ell_\g(\varphi^\infty_\ns)}-\frac{|\upsilon_\s|s_\l R^2(t)}{3D_\g}\right\},
\end{equation}
Since no reactions occur in the absence of material, $s(\bar0)=0$, we can expect at least $[-s_\g]/[-\bar k_\g^\T\bar\upsilon]=\mathcal{O}(\varphi)$, meaning that the acceleration due to gas-phase reactions is subleading\,---\,the second term is smaller than the first one by $R/\ell_\g\ll1$. The last term, originating from dense-phase reactions, competes with the passive diffusive growth, and it can prevent it entirely once all droplets reach the stable size-control radius
\begin{equation}
    R_\mathrm{sc}=\sqrt{\frac{3D_\g}{|\upsilon_\s|s_\l}[\varphi_\s^\infty-\phi_{\s\g}^\L(R_\mathrm{sc})]}.
\end{equation}
Note that as droplets grow, the supersaturated dilute phase loses material, meaning that $\varphi_\s^\infty$ is indirectly also affected by $R_\mathrm{sc}$. Thus, to avoid size control, we need to transition to the next regime where the dense-phase flux reduces to $R^2J_\l\sim s_\l R^2$ (for $R\gg\ell_\l$) prior to reaching $R_\mathrm{sc}$. This translates to the requirement $R_\mathrm{sc}\gg\ell_\l$, or
\begin{equation}
    \frac{\ell_\l^2s_\l}{D_\g}\ll \varphi_\s^\infty-\phi_{\s\g}^\L(\ell_\l),
\end{equation}
shown in the main text.

\subsubsection{Large droplet}\label{sec:large_droplet}

This time we assume $\ell_\g,\ell_\l\ll R$. Using Eq.~\eqref{eq:Ginterpolant}, Eq.~\eqref{eq:radius_growth_phi_nsg} becomes
\begin{equation}
    \frac{\ell_\g(\phi_{\ns\g}^\L)}R(\varphi^\infty_\ns-\phi_{\ns\g}^\L)=-\frac{\ell_\g^2(\phi_{\ns\g}^\L)}{D_\g}\left[ s_\g(\phi_{\ns\g}^\L)+s_\l\frac{\ell_\l}{\ell_\g(\phi_{\ns\g}^\L)}\left(1-\frac{\ell_\l}R\right)\right],\label{eq:radius_growth_phi_nsg_large}
\end{equation}
Here, without further assumptions (as in Sec.~\ref{sec:gas_example} and in the main text) $\phi_{\ns\g}^\L=\phi_{\ns\g}^\circ+\lambda_{\ns\g}/R$ can only be found implicitly. To leading-order in $\ell_\g/R,\lambda_{\ns\g}/R\ll1$, $\phi_{\ns\g}^\circ$ must satisfy implicitly
\begin{equation}
   -s_\g(\phi_{\ns\g}^\circ)\ell_\g(\phi_{\ns\g}^\circ)=s_\l\ell_\l.\label{eq:phi_nsg_large0}
\end{equation}
That is, since the reactions occur with rates proportional to the droplet surface, while the droplet volume can only grow at a rate proportional to the radius due to the mass conservation constraint (Eq.~\eqref{eq:radius_growth_conserved}), the reactions must balance each other to leading order for a large droplet, $(4\pi R^2\ell_\g)s_\g+(4\pi R^2\ell_\l)s_\l=0$, where the corresponding reactive shell widths around the interface are $\ell_\alpha$ ($\alpha=\l,\g$).

To find the next-order, we Taylor-expand $s_\g(\phi_{\ns\g}^\L)$ and $\ell_\g(\phi_{\ns\g}^\L)$ around $\phi_{\ns\g}^\circ$. For brevity, we will show the result for a slightly simpler case where $\ell_\g=\mathrm{const}$. We prepare $s_\g(\phi_{\ns\g}^\L)=s_\g(\phi_{\ns\g}^\circ)+(\partial s/\partial\phi_{\ns})|_{\phi_{\ns\g}^\circ}\lambda_{\ns\g}/R=-s_\l\ell_\l/\ell_\g+k_{\ns}(\phi_{\ns\g}^\circ)\lambda_{\ns\g}/R$, where we used Eq.~\eqref{eq:phi_nsg_large0} and identified the first-order coefficient $k_{\ns}$ (Eq.~\eqref{eq:first}). Thence, inserting this expansion in Eq.~\eqref{eq:radius_growth_phi_nsg_large}, $\lambda_{\ns\g}$ reads
\begin{equation}
    \frac{\lambda_{\ns\g}}R=\frac{D_\g}{-k_{\ns}(\phi_{\ns\g}^\circ)\ell_\g^2}\left(\varphi^\infty_\ns-\phi_{\ns\g}^\circ-\frac{s_\l\ell_\l^2}{D_\g}\right)\frac{\ell_\g}R.\label{eq:lambda_nsg_big}
\end{equation}

Combining, we find
\begin{equation}
    \frac{\d R(t)}{\d t}=\frac{D_\g}{R(t)}\left[(\varphi^\infty-\phi_{\s\g}^\circ-|\upsilon_\s|\phi_{\ns\g}^\circ)-\frac{\lambda_{\s\g}}{R(t)}\left(1-\frac{s_\l\ell_\l^2}{[-k_{\ns}(\phi_{\ns\g}^\circ)]\ell_\g^2}\frac{\ell_\g}{\lambda_{\s\g}}+\frac{D_\g(\varphi_{\ns}^\infty-\phi_{\ns\g}^\circ)}{[-k_{\ns}(\phi_{\ns\g}^\circ)]\ell_\g\lambda_{\s\g}}\right)\right].
\end{equation}
To make progress, we consider the gas-phase reaction mechanism considered in the text, $s(\bar\phi_\g)=k_{\s\to\ns}\phi_{\s\g}-k_{\ns\to\s}\phi_{\ns\g}$ with $\upsilon_\s=-1$. In this case, $-s(\phi_{\ns\g}^\circ)=k_{\ns\to\s}\phi_{\ns\g}^\circ-k_{\s\to\ns}(\phi_{\s\g}^\circ+\lambda_{\s\g}/R)$, $\ell_\g^2=D_\g/(k_{\s\to\ns}+k_{\ns\to\s})$, and $- k_\ns=k_{\ns\to\s}$. Consequently, we have $\phi_{\ns\g}^\circ=(k_{\s\to\ns}/k_{\ns\to\s})(\phi_{\s\g}^\circ+\lambda_{\s\g}/R)+(s_\l/k_{\ns\to\s})(\ell_\l/\ell_\g)$ and $\varphi_{\ns}^\infty=k_{\s\to\ns}\varphi^\infty/(k_{\s\to\ns}+k_{\ns\to\s})$, so
\begin{equation}
    \frac{\d R(t)}{\d t}=\frac{D_\mathrm{eff}}{R(t)}\left[(\varphi_\mathrm{eff}^\infty-\phi_{\s\g}^\circ)-\frac{\lambda_\mathrm{eff}}{R(t)}\right],
\end{equation}
where
\begin{eqnarray}
    \frac{D_\mathrm{eff}}{D_\g}&=&1+\frac{k_{\s\to\ns}}{k_{\ns\to\s}},\\
    \varphi_\mathrm{eff}^\infty&=&\frac{k_{\ns\to\s}\varphi^\infty-s_\l(\ell_\l/\ell_\g)}{k_{\s\to\ns}+k_{\ns\to\s}},\\
    \frac{\lambda_\mathrm{eff}}{\lambda_{\s\g}}&=&1-\frac{s_\l\ell_\l^2}{D_\g}\left(1+\frac{\ell_\g}{\ell_\l}\right)\frac{\ell_\g}{\lambda_{\s\g}}+\frac{k_{\s\to\ns}}{k_{\ns\to\s}}\frac{\ell_\g}{\lambda_{\s\g}}(\varphi_\mathrm{eff}^\infty-\phi_{\s\g}^\circ).\label{eq:dRdtlambda}
\end{eqnarray}

When the reactions inside the droplet are sufficiently fast, $s_\l\ell_\l^2/D_\g\gg(1+\ell_\g/\ell_\l)(\lambda_{\s\g}/\ell_\g)$, we may neglect the contribution from the Laplace pressure, the first term $1$ in Eq.~\eqref{eq:dRdtlambda}. Thus, temporarily ignoring the third term in Eq.~\eqref{eq:dRdtlambda}, the reactions cause an effective negative Laplace pressure correction coefficient, $\lambda_\mathrm{eff}=-s_\l\ell_\l\ell_\g(\ell_\l+\ell_\g)/D_\g$. As a result, the droplets do not undergo ripening \textit{per se}, i.e., smaller droplets do not shrink. Instead, all droplets that pass the $R\gg\ell_\l,\ell_\g$ threshold are expected to grow until the extensive, stable radius $R_\mathrm{nr}=\lambda_\mathrm{eff}/(\varphi_\mathrm{eff}^\infty-\phi_{\s\g}^\circ)\sim V^{1/3}$ is reached. For this stable radius, the effective supersaturation $\varphi_\mathrm{eff}^\infty-\phi_{\s\g}^\circ$ becomes negative. Thus, now including the third term of Eq.~\eqref{eq:dRdtlambda}, we see that it will further contribute to the already negative $\lambda_\mathrm{eff}$, and overall
\begin{equation}
    \frac{\d R(t)}{\d t}=\frac{D_\mathrm{eff}}{R(t)}\left[-|\varphi_\mathrm{eff}^\infty-\phi_{\s\g}^\circ|+\frac{|\lambda_\mathrm{eff}|}{R(t)}\right].
\end{equation}
As growth continues, material is being taken up from the dilute phase, so $\varphi_\mathrm{eff}^\infty-\phi_{\s\g}^\circ$ becomes even more negative so $|\varphi_\mathrm{eff}^\infty-\phi_{\s\g}^\circ|$ increases. This ripening arrest has been recently reported in Ref.~\cite{BauermannARXIV24}; we will not investigate it further here.

We now discuss the sought-after regime where Ostwald ripening is accelerated. For slow reactions inside the droplet, $s_\l\ell_\l^2/D_\g\ll(1+\ell_\g/\ell_\l)(\lambda_{\s\g}/\ell_\g)$, the Laplace pressure dominates as the second term of Eq.~\eqref{eq:dRdtlambda} is negligible. Thus, the Laplace pressure correction coefficient is positive, and ripening occurs as expected. Here, $\lambda_\mathrm{eff}=\lambda_{\s\g}+(k_{\s\to\ns}/k_{\ns\to\s})\ell_\g(\varphi_\mathrm{eff}^\infty-\phi_{\s\g}^\circ)$, where the effective supersaturation $\varphi_\mathrm{eff}^\infty-\phi_{\s\g}^\circ$, as in standard Ostwald ripening, is positive. Since $\ell_\g/\lambda_{\s\g}$ is a large number (reactions occur over lengthscales larger than the interface width), the contribution from the small $\varphi_\mathrm{eff}^\infty-\phi_{\s\g}^\circ$ may be transiently important. During a transient where $\lambda_{\s\g}\ll(k_{\s\to\ns}/k_{\ns\to\s})\ell_\g(\varphi_\mathrm{eff}^\infty-\phi_{\s\g}^\circ)$, we find 
\begin{equation}
    \frac{\d R}{\d t}=\frac{D_\mathrm{eff}(\varphi_\mathrm{eff}^\infty-\phi_{\s\g}^\circ)}{R}\left(1+\frac{k_{\s\to\ns}}{k_{\ns\to\s}}\frac{\ell}{R}\right)\simeq \frac{D_\mathrm{eff}(\varphi_\mathrm{eff}^\infty-\phi_{\s\g}^\circ)}{R},
\end{equation}
meaning that $R^2\sim t$ until the effective supersaturation has lowered sufficiently, and the asymptotic $\lambda_{\s\g}\gg(k_{\s\to\ns}/k_{\ns\to\s})\ell_\g(\varphi_\mathrm{eff}^\infty-\phi_{\s\g}^\circ)$ regime is reached.
Once $\varphi_\mathrm{eff}^\infty-\phi_{\s\g}^\circ$ decreased sufficiently, we reproduce the standard $\lambda_\mathrm{eff}=\lambda_{\s\g}$, and hence
\begin{equation}
    \frac{\d R(t)}{\d t}=\frac{D_\mathrm{eff}}{R(t)}\left[(\varphi_\mathrm{eff}^\infty-\phi_{\s\g}^\circ)-\frac{\lambda_{\s\g}}{R(t)}\right],\label{eq:dRdt_accel}
\end{equation}
which is Eq.~(9) of the main text.

\subsection{Lifshitz-Slyozov scaling}

We conclude by demonstrating the Lifshitz-Slyozov scaling. The total amount of material (sticky and nonsticky) is conserved, thus implicitly connecting the saturation $\varphi^\infty$ to the radii of all droplets present in the solution. Namely, defining the droplet-radius distribution $f(R,t)$ (normalized such that $V\int_0^\infty \d Rf(R,t)$ is the total number of droplets in the container), material conservation across the sparse droplets and their dilute surroundings can be written as
\begin{equation}
    \varphi^\infty(t)V+(\phi_{\s\l}^\circ+\phi_{\ns\l}^\circ-\varphi^\infty(t))\int_0^\infty\d R[Vf(R,t)] \frac43\pi R^3=\mathrm{const},\label{eq:conserve}
\end{equation}
where the second term is the increase in volume fraction of the total amount of material due to forming dense droplets. Note that the total individual amounts of sticky and nonsticky particles is not conserved due to reactions; only their combined amount. In Eq.~\eqref{eq:conserve}, we ignored the Laplace-pressure corrections to the liquid-phase densities to leading order (the droplets are large), and considered a uniform dilute-phase volume fraction thus ignoring the crossover from $\phi_{\s\g}^\circ+\phi_{\ns\g}^\circ$ to $\varphi^\infty$ occurring over a distance $\ell_\g$ around each droplet (which is assumed much smaller than the inter-droplet distance). 

To bring it to the Lifshitz-Slyozov form, we further neglect $\phi_{\ns\l}^\circ,\varphi^\infty\ll\phi_{\s\l}^\circ$, and by multiplying Eq.~\eqref{eq:conserve} by the constant $k_\mathrm{ns\to\s}/(k_{\s\to\ns}+k_{\ns\to\s})$ and subtracting the constants $s_\l(\ell_\l/\ell_\g)/(k_{\s\to\ns}+k_{\ns\to\s})$ and $\phi_{\s\g}^\circ$, we find
\begin{equation}
    (\varphi_\mathrm{eff}^\infty-\phi_{\s\g}^\circ)+\frac{k_\mathrm{ns\to\s}}{k_{\s\to\ns}+k_{\ns\to\s}}\phi_{\s\l}^\circ\int_0^\infty\d Rf(R,t) \frac43\pi R^3=\mathrm{const}.\label{eq:conserve_eff}
\end{equation}
Finally, redefining $k_\mathrm{ns\to\s}f(R,t)/(k_{\s\to\ns}+k_{\ns\to\s})\to f(R,t)$, we reach the starting point of Lifshitz-Slyozov analysis, Eqs.~\eqref{eq:dRdt_accel} and~\eqref{eq:conserve_eff}, but with effective supersaturation and transport coefficients.

As droplets grow, they take up sticky particles from the supersaturated gas phase, meaning that the saturation $\varphi_\mathrm{eff}^\infty$ decreases over time. Therefore, according to Eq.~\eqref{eq:dRdt_accel}, droplets that were previously able to grow may become unstable to Laplace pressure and will now shrink, while the larger droplets continue growing. (This process continues until a single large droplet is left such that $\varphi_\mathrm{eff}^\infty=\phi_{\s\g}^\circ$.) Thus, the canonical Lifshitz-Slyozov scaling law emerges~\cite{LifshitzJPCS61}: The overall number of droplets decreases in time (as the smaller droplets vanish; note that the droplet number distribution was rescaled by the above factor), the saturation decreases too, while the average droplet radius grows as
\begin{equation}
    \langle R(t)\rangle^3-\langle R(0)\rangle^3_\mathrm{a}=\frac49D_\mathrm{eff}\lambda_{\s\g} t.\label{eq:passive_ripening}
\end{equation}
As mentioned in the main text, the accelerated regime occurs after multiple activity-related transients, meaning that the intercept $\langle R(0)\rangle^3_\mathrm{a}$ might differ from the passive case.

\end{widetext}

\end{document}